\documentclass[11pt,a4paper]{article}

\def\DottedCircle{
\qbezier[4](0.966,-0.259)(1.04,0)(0.966,0.259)
\qbezier[4](0.966,0.259)(0.897,0.518)(0.707,0.707)
\qbezier[4](0.707,0.707)(0.518,0.897)(0.259,0.966)
\qbezier[4](0.259,0.966)(0,1.04)(-0.259,0.966)
\qbezier[4](-0.259,0.966)(-0.518,0.897)(-0.707,0.707)
\qbezier[4](-0.707,0.707)(-0.897,0.518)(-0.966,0.259)
\qbezier[4](-0.966,0.259)(-1.04,0)(-0.966,-0.259)
\qbezier[4](-0.966,-0.259)(-0.897,-0.518)(-0.707,-0.707)
\qbezier[4](-0.707,-0.707)(-0.518,-0.897)(-0.259,-0.966)
\qbezier[4](-0.259,-0.966)(0,-1.04)(0.259,-0.966)
\qbezier[4](0.259,-0.966)(0.518,-0.897)(0.707,-0.707)
\qbezier[4](0.707,-0.707)(0.897,-0.518)(0.966,-0.259)
}
\def\FullCircle{
\thicklines
\put(0,0){\circle{2}}
}
\def\Endpoint[#1]{
\ifcase#1
\put(1,0){\circle*{0.15}}
\or\put(0.866,0.5){\circle*{0.15}}
\or\put(0.5,0.866){\circle*{0.15}}
\or\put(0,1){\circle*{0.15}}
\or\put(-0.5,0.866){\circle*{0.15}}
\or\put(-0.866,0.5){\circle*{0.15}}
\or\put(-1,0){\circle*{0.15}}
\or\put(-0.866,-0.5){\circle*{0.15}}
\or\put(-0.5,-0.866){\circle*{0.15}}
\or\put(0,-1){\circle*{0.15}}
\or\put(0.5,-0.866){\circle*{0.15}}
\or\put(0.866,-0.5){\circle*{0.15}}
\fi}
\def\Arc[#1]{
\thicklines			
\ifcase#1
\qbezier[25](0.966,-0.259)(1.04,0)(0.966,0.259)
\or
\qbezier[25](0.966,0.259)(0.897,0.518)(0.707,0.707)
\or
\qbezier[25](0.707,0.707)(0.518,0.897)(0.259,0.966)
\or
\qbezier[25](0.259,0.966)(0,1.04)(-0.259,0.966)
\or
\qbezier[25](-0.259,0.966)(-0.518,0.897)(-0.707,0.707)
\or
\qbezier[25](-0.707,0.707)(-0.897,0linde.518)(-0.966,0.259)
\or
\qbezier[25](-0.966,0.259)(-1.04,0)(-0.966,-0.259)
\or
\qbezier[25](-0.966,-0.259)(-0.897,-0.518)(-0.707,-0.707)
\orY-Song. Piao, Phys. Rev. {\bf D 74}, 047302 (2006);
\qbezier[25](-0.707,-0.707)(-0.518,-0.897)(-0.259,-0.966)
\or
\qbezier[25](-0.259,-0.966)(0,-1.04)(0.259,-0.966)
\or
\qbezier[25](0.259,-0.966)(0.518,-0.897)(0.707,-0.707)
\or
\qbezier[25](0.707,-0.707)(0.897,-0.518)(0.966,-0.259)
\fi}
\def\DottedArc[#1]{
\ifcase#1
\qbezier[4](0.966,-0.259)(1.04,0)(0.966,0.259)
\or
\qbezier[4](0.966,0.259)(0.897,0.518)(0.707,0.707)
\or
\qbezier[4](0.707,0.707)(0.518,0.897)(0.259,0.966)
\or
\qbezier[4](0.259,0.966)(0,1.04)(-0.259,0.966)
\or
\qbezier[4](-0.259,0.966)(-0.518,0.897)(-0.707,0.707)
\or
\qbezier[4](-0.707,0.707)(-0.897,0.518)(-0.966,0.259)
\or
\qbezier[4](-0.966,0.259)(-1.04,0)(-0.966,-0.259)
\or
\qbezier[4](-0.966,-0.259)(-0.897,-0.518)(-0.707,-0.707)
\or
\qbezier[4](-0.707,-0.707)(-0.518,-0.897)(-0.259,-0.966)
\or
\qbezier[4](-0.259,-0.966)(0,-1.04)(0.259,-0.966)
\or
\qbezier[4](0.259,-0.966)(0.518,-0.897)(0.707,-0.707)
\or
\qbezier[4](0.707,-0.707)(0.897,-0.518)(0.966,-0.259)
\fi}
\def\Chord[#1,#2]{
\thinlines
\ifnum#1>#2\Chord[#2,#1]
\else\ifnum#1<#2
\ifcase#1
\ifcase#2
\or\qbezier(1,0)(0.516,0.138)(0.866,0.5)
\or\qbezier(1,0)(0.45,0.26)(0.5,0.866)
\or\qbezier(1,0)(0.327,0.327)(0,1)
\or\qbezier(1,0)(0.179,0.311)(-0.5,0.866)
\or\qbezier(1,0)(0.0536,0.2)(-0.866,0.5)
\or\put(1, 0){\line(-2, 0){2}}
\or\qbezier(1,0)(0.0536,-0.2)(-0.866,-0.5)
\or\qbezier(1,0)(0.179,-0.311)(-0.5,-0.866)
\or\qbezier(1,0)(0.327,-0.327)(0,-1)
\or\qbezier(1,0)(0.45,-0.26)(0.5,-0.866)
\or\qbezier(1,0)(0.516,-0.138)(0.866,-0.5)
\fi
\or\ifcase#2\or
\or\qbezier(0.866,0.5)(0.378,0.378)(0.5,0.866)
\or\qbezier(0.866,0.5)(0.26,0.45)(0,1)
\or\qbezier(0.866,0.5)(0.12,0.446)(-0.5,0.866)
\or\qbezier(0.866,0.5)(0,0.359)(-0.866,0.5)
\or\qbezier(0.866,0.5)(-0.0536,0.2)(-1,0)
\or\put(0.866, 0.5){\line(-5, -3){1.73}}
\or\qbezier(0.866,0.5)(0.146,-0.146)(-0.5,-0.866)
\or\qbezier(0.866,0.5)(0.311,-0.179)(0,-1)
\or\qbezier(0.866,0.5)(0.446,-0.12)(0.5,-0.866)
\or\qbezier(0.866,0.5)(0.52,0)(0.866,-0.5)
\fi
\or\ifcase#2\or\or
\or\qbezier(0.5,0.866)(0.138,0.516)(0,1)
\or\qbezier(0.5,0.866)(0,0.52)(-0.5,0.866)
\or\qbezier(0.5,0.866)(-0.12,0.446)(-0.866,0.5)
\or\qbezier(0.5,0.866)(-0.179,0.311)(-1,0)
\or\qbezier(0.5,0.866)(-0.146,0.146)(-0.866,-0.5)
\or\put(0.5, 0.866){\line(-3, -5){1}}
\or\qbezier(0.5,0.866)(0.2,-0.0536)(0,-1)
\or\qbezier(0.5,0.866)(0.359,0)(0.5,-0.866)
\or\qbezier(0.5,0.866)(0.446,0.12)(0.866,-0.5)
\fi
\or\ifcase#2\or\or\or
\or\qbezier(0,1.)(-0.138,0.516)(-0.5,0.866)
\or\qbezier(0,1.)(-0.26,0.45)(-0.866,0.5)
\or\qbezier(0,1.)(-0.327,0.327)(-1,0)
\or\qbezier(0,1.)(-0.311,0.179)(-0.866,-0.5)
\or\qbezier(0,1.)(-0.2,0.0536)(-0.5,-0.866)
\or\put(0, 1){\line(0, -2){2}}
\or\qbezier(0,1.)(0.2,0.0536)(0.5,-0.866)
\or\qbezier(0,1.)(0.311,0.179)(0.866,-0.5)
\fi
\or\ifcase#2\or\or\or\or
\or\qbezier(-0.5,0.866)(-0.378,0.378)(-0.866,0.5)
\or\qbezier(-0.5,0.866)(-0.45,0.26)(-1,0)
\or\qbezier(-0.5,0.866)(-0.446,0.12)(-0.866,-0.5)
\or\qbezier(-0.5,0.866)(-0.359,0)(-0.5,-0.866)
\or\qbezier(-0.5,0.866)(-0.2,-0.0536)(0,-1)
\or\put(-0.5, 0.866){\line(3, -5){1}}
\or\qbezier(-0.5,0.866)(0.146,0.146)(0.866,-0.5)
\fi
\or\ifcase#2\or\or\or\or\or
\or\qbezier(-0.866,0.5)(-0.516,0.138)(-1,0)
\or\qbezier(-0.866,0.5)(-0.52,0)(-0.866,-0.5)
\or\qbezier(-0.866,0.5)(-0.446,-0.12)(-0.5,-0.866)
\or\qbezier(-0.866,0.5)(-0.311,-0.179)(0,-1)
\or\qbezier(-0.866,0.5)(-0.146,-0.146)(0.5,-0.866)
\or\put(-0.866, 0.5){\line(5, -3){1.73}}
\fi
\or\ifcase#2\or\or\or\or\or\or
\or\qbezier(-1,0)(-0.516,-0.138)(-0.866,-0.5)
\or\qbezier(-1,0)(-0.45,-0.26)(-0.5,-0.866)
\or\qbezier(-1,0)(-0.327,-0.327)(0,-1)
\or\qbezier(-1,0)(-0.179,-0.311)(0.5,-0.866)
\or\qbezier(-1,0)(-0.0536,-0.2)(0.866,-0.5)
\fi
\or\ifcase#2\or\or\or\or\or\or\or
\or\qbezier(-0.866,-0.5)(-0.378,-0.378)(-0.5,-0.866)
\or\qbezier(-0.866,-0.5)(-0.26,-0.45)(0,-1)
\or\qbezier(-0.866,-0.5)(-0.12,-0.446)(0.5,-0.866)
\or\qbezier(-0.866,-0.5)(0,-0.359)(0.866,-0.5)
\fi
\or\ifcase#2\or\or\or\or\or\or\or\or
\or\qbezier(-0.5,-0.866)(-0.138,-0.516)(0,-1)
\or\qbezier(-0.5,-0.866)(0,-0.52)(0.5,-0.866)
\or\qbezier(-0.5,-0.866)(0.12,-0.446)(0.866,-0.5)
\fi
\or\ifcase#2\or\or\or\or\or\or\or\or\or
\or\qbezier(0,-1.)(0.138,-0.516)(0.5,-0.866)
\or\qbezier(0,-1.)(0.26,-0.45)(0.866,-0.5)
\fi
\or\ifcase#2\or\or\or\or\or\or\or\or\or\or
\or\qbezier(0.5,-0.866)(0.378,-0.378)(0.866,-0.5)
\fi\fi\fi\fi}
\def\FullChord[#1,#2]{
\Endpoint[#1]
\Endpoint[#2]
\Arc[#1]
\Arc[#2]
\Chord[#1,#2]
}
\def\EndChord[#1,#2]{
\Endpoint[#1]
\Endpoint[#2]
\Chord[#1,#2]
}
\def\Picture#1{
\begin{picture}(2,1)(-1,-0.167)
#1
\end{picture}
}
\def\DottedChordDiagram[#1,#2]{
\Picture{\DottedCircle \FullChord[#1,#2]}
}
\usepackage{amsmath,amsthm,amssymb}
\usepackage{graphicx}
\usepackage{epsfig,cite}
\usepackage[dvips]{color}
\usepackage{multicol}
\makeatletter
\@addtoreset{equation}{section}
\makeatother

\newcommand{\abb}{\addtolength{\belowdisplayskip}{\belowdisplayskip}}

\newcommand{\Tr}{{\rm Tr}}
\newcommand{\ssk}{{\hspace{3pt}}}

\newcommand{\bp}{  \begin{pmatrix} }
\newcommand{\ep}{  \end{pmatrix} }

\begin{document}
\titlepage

\begin{flushright}
\end{flushright}
\vspace{1cm}
\begin{center}
\Large\textbf{DBI N-flation}\\
\vspace{2cm} 
\small\textbf{John Ward}\footnote{jwa@uvic.ca}\\
\end{center}
\begin{center}
\emph{Department of Physics and Astronomy, University of Victoria, Victoria \\
BC, V8P 1A1, Canada}
\end{center}
\begin{abstract}
We investigate DBI inflation using $N$ multiple branes and show how the configuration is equivalent to a single wrapped $D5$-brane with 
flux. We then show how $1/N$ corrections can be implemented, and we examine the sound speed and levels of non-Gaussianities
in two distinct cases. For models with constant warping we find that the non-Gaussian amplitude is bounded from above (as a function of $\gamma$).
For $AdS$ backgrounds we find that the signature is generally large and positive, although is no longer globally defined over the full
phase space. We then investigate an inflationary mechanism using a representation cascade, whereby the transition from a reducible
representation to the irrep drives inflation. 
\end{abstract}

\newpage
\section{Introduction}
In the absence of any direct test of string theory, cosmology remains the best laboratory with which to test string theoretic models \cite{lectures}.
Over the past few years we have witnessed cosmology become a precision science, with COBE, WMAP and SDSS \cite{data} providing crucial support
for the flatness of the universe, the existence of dark energy and for a period of cosmic inflation. Whilst the dark energy puzzle
remains an outstanding problem for theoretical physics in general \cite{darkenergy}, inflation has been a carefully developed paradigm with many explicit
models. Unfortunately as far as inflationary model building is concerned, there are still many problems to be resolved. Particularly
since many of the models suffer from super-Planckian VEV's for the inflaton field \cite{lectures}, and therefore find themselves in a region where 
quantum gravity effects are non-negligible. Conversely the lack of a background independent formulation of string theory has prevented
the explicit construction of top-down models, and much of string cosmology has been done explicitly at the field theory level. Whilst there
is nothing wrong with this in principle, many of the models are somehow missing much of the underlying string theoretic structure which
is where we would expect the more interesting physics to emerge.

There are indications that this picture is about to change \cite{lectures}. Our understanding of both geometric \cite{geometricflux} and non-geometric flux compactifications 
\cite{nongeomflux} of type II string theory has increased immeasurably in recent years, allowing for the construction of more realistic inflationary
models \cite{realisticmodels}. 
Additionally models emerging from heterotic M-theory \cite{assistedinflationmtheory} can also now be placed on a more secure footing, and may yet unify both the 
standard model and inflaton sectors.  Of course there remains much work to be done, but the general prognosis is that inflationary model
building will only improve.

One of the simpler models of string inflation relies on the motion of branes, 
where the inflaton is now reflected in terms of geometry. Either
as the distance between a pair of $D3-\bar{D}3$-branes \cite{dbarinflation}, or as the distance between a single brane and some reference point in a warped throat 
\cite{braneinflation}. These
models are especially appealing, not just because of their simplicity, but because the inflaton is an open string mode which will vanish
at the end of inflation and therefore one doesn't need to worry about how it interacts with the standard model sector \cite{reheating}. In the light of 
recent developments in type IIB flux compactifications, and the existence of a potential multitude of warped throats with which
to resolve the hierarchy problem, these models have become even more appealing. Given the vast number of free parameters that we often find in 
string model building, it is relatively 
easy to construct a model that satisfies the WMAP data. Therefore we should be interested in predictions that can be ruled out. These should
not be regarded as being deficiencies of string theory, on the contrary in fact, as we are narrowing the parameter space with each
one eliminated. Much of the community is now involved in determining which signatures of a particular model can be tested. Indeed many of them
rely on bounds placed on cosmic (super)string formation during or after inflation \cite{cosmicstrings}.

One particular model based on the non-linear structure of the DBI action itself, named DBI inflation \cite{dbiinflation}, has an interesting
signature in that it predicts large levels of non-Gaussian perturbations during inflation \cite{nongaussianities}. 
This is important since the result is apparently background independent \cite{lectures}. 
There has been much work on this model and its implications \cite{irinflation, related, spinflation, slowroll}, 
but the general
consensus now is that the simplest scenario is no longer viable. This means it is essential for us to develop more realistic variations of
this model\cite{assistedinflationdbi, multibrane, ward}. 
There have been several proposals for extending this work, ranging from 
multi-brane configurations to branes wrapping non-trivial cycles. Although these extensions are able to satisfy the experimental bounds, 
there is still some concern about the range of validity of such models.

In this paper we will initiate an investigation into the multi-brane proposal of \cite{ward}, and begin to institute higher order corrections to the 
action \cite{finiten}. In this case the corrections we are interested in are the $1/N$ corrections in the large $N$ limit. These corrections
are important as the primary constraint on the model is that $N<<M$, where $N$ is the number of branes and $M$ is the total flux in a throat.
In the compact case we know that $M$ is bounded from above, by considering compactification of F-theory onto Calabi-Yau (CY) four-folds \cite{geometricflux}, 
and therefore this restricts the number of dynamical branes accordingly. The large $N$ limit ensures that the action simplifies, moreover
the relevant physical scales are either suppressed or enhanced by this large number allowing us to evade many of the tight constraints. However 
assuming large $N$ also means that the backreaction could be dangerously out of control. What we would like is therefore to keep $N$ relatively 
large but also understand how some of the $1/N$ corrections alter things.
Given the highly non-linear structure of the DBI even at leading order, we expect that these corrections will be analytically complicated. Therefore
rather than search for concrete models of inflation we will restrict ourselves to analysis of the interesting observables associated with DBI inflation.
In short we will be interested in i) how the corrections alter the sound speed and ii) how the corrections alter the prediction of large non-Gaussianity.
We leave a more detailed investigation of the model dynamics to another publication.

In section II we will introduce the action for the multi-brane configuration we are describing and its features. We will also demonstrate how this is
equivalent to a model based on a $D5$-brane wrapping a two-cycle and carrying flux along its internal directions - and therefore overlapping
with the model proposed in \cite{wrappedbranes}. In section III we will show how these $1/N$ corrections can be implemented, and how they alter the leading order behaviour
of the Lagrangian. We will then investigate how the corrections alter the predictivity of the model. In the final section we will consider an 
alternative model using a similar set-up, but we will use the group representation space as the inflationary phase space.

\section{Multiple brane inflation}
Our primary interest here is to study DBI inflation driven by multiple branes in a warped geometry \cite{assistedinflationdbi, multibrane, ward, giantinflaton}. 
By now there is a considerable mass of evidence to suggest that the
simplest $D3$-brane scenario does not lead to new physically observable signatures, and is therefore indistinguishable from standard slow roll inflaton models \cite{constraints, wrappedbranes}.
As a result we must beyond the simplest models, and search for other regions of solution space which could give us inflationary trajectories. Perhaps the
next simplest approach, which we will consider here, is to replace the single brane by $N$ $D3$-branes and study the corresponding dynamics. In particular
we will consider the case where all the branes are localised at distances less than the string length. In terms of the world-volume field theories this means
that we are studying the $U(N)$ theory rather than the $U(1)^N$ theory \cite{myers, tseytlin}. This differs significantly from a theory of $N$-branes that are 
separated at larger distances \cite{assistedinflationdbi, multibrane}, which will fall into the class of Assisted Inflation \cite{assistedinflation}. 
What is important in these models is that each of the $N$-fields
follow an attractor trajectory so that we can treat each field as having an equal contribution to the Hubble scale. If they do not, then we will
generally find signatures of isocurvature modes. Although these were shown to be suppressed in the case of DBI inflation \cite{assistedinflationdbi}. Other models using the large $N$ limit typically involve axions \cite{nflation, kallosh}.

Unlike in the case of a single $D$-brane, the action for multiple coincident branes is still unknown. As a leading order solution we will employ the use of the
Myers action, which is known to deviate from the full string theory scattering amplitude at $\mathcal{O}(F^6)$ \cite{myers}. However despite this not being the full solution,
it will almost certainly be part of the full solution - and therefore one should regard our model building program as being the first step towards the complete string description.

One may also regard this solution as being more generic than the single brane models - since we expect these objects to be created quantum mechanically 
at the end of brane/flux annihilation \cite{giantinflaton}. Tuning the fluxes to ensure that only a single $D3$-brane emerges through this process imposes additional fine tuning of the
parameters, and is often unsatisfactory. Moreover the annihilation itself is reasonably well understood, and the residual branes will find themselves localised
in the IR end of a warped throat \cite{irinflation}. The relevant physical scales are thus significantly red-shifted with regard to an observer sitting in the bulk space, and one
hope is that the standard model will be localised upon some intersecting brane stack in another throat.

We will take the ten-dimensional background metric to be of the following form, which can be regarded as a cone over the base space $X_5$
\begin{equation}
ds_{10}^2 = h^2 g_{\mu \nu}dx^{\mu} dx^{\nu} + h^{-2} (d\rho^2 + \rho^2 dX_5^2)
\end{equation}
where the radial direction is parameterised by $\rho$ rather than $r$ since the latter is often assigned to the ratio of tensor to scalar perturbations.
The factors of $h$ are the warp factors for the geometry, and are functions of the transverse coordinates.
For simplicity we will set all gauge fields to zero, and in addition we will assume that the NS two-form $B^{(2)}$ is also zero\footnote{In the Klebanov-Strassler geometry \cite{klebanovstrassler}
the two-form runs logarithmically with the radial displacement, and so can be tuned to vanish near the tip of the throat.} since this simplifies things considerably.
We assume that the warped inflationary throat is one of many throats glued onto the internal Calabi-Yau space, although the gluing is a model dependent effect and
may well induce corrections to the flux induced potential\cite{geometricflux}
We will assume that the Chern-Simons sector consists solely of the RR four-form $C^{(4)}$, and that this is simply proportional to the warp factor.

The relevant contribution to the Myers action \cite{myers} can be written as follows
\begin{equation}
S=-T_p \int d^{p+1} \xi \rm{STr} \left( \sqrt{-\rm{det}(\hat{E}_{\mu \nu}+\hat{E}_{\mu i}(Q^{-1}-\delta)^{ij} \hat{E}_{j\nu})}\sqrt{\rm{det} Q^i_j} \right)
\end{equation}
supplemented by the non-Abelian Chern-Simons contribution
\begin{equation}
S_{CS} = \mu_p \int d^{p+1} \xi \rm{STr}\left(e^{i \lambda i_{\phi} i_{\phi}} \sum \hat{C}^{(n)} \right).
\end{equation}
Let us explain the terminology used above. Firstly the scalar fields are now matrix valued, and therefore we require a prescription for taking their trace. This is done using the symmetrised trace, which requires us to take the fully symmetric averaging over all possible orderings before taking the trace. This is 
required in order to reproduce the known (lowest order) string scattering amplitudes. The kinetic term in the action contains the matrix $Q^i_j$, which also appears as a potential term. This matrix is explicitly given by
\begin{equation}
Q^i_j + \delta^i_j + i \lambda [\phi^i, \phi^j]E_{kj}
\end{equation}
and we will work to leading order in its expansion. The metric $E_{\mu \nu}$ is a linear combination of the metric
and NS two-form, although in our simplistic scenario it reduces to the metric only. Greek indices run over the non-compact
directions, whilst roman indices correspond to the transverse directions.
As usual hats denote pullbacks of the space time fields to the world-volume. The Chern-Simons term involves a summation
over all possible RR fields present in the theory, coupled to an expansion of so-called interior derivatives.
Usually the Chern-Simons term involves coupling to lower degree form fields through the introduction of non-vanishing
Chern classes, the introduction of the interior derivatives also induces couplings to higher dimensional form
fields through their action
\begin{equation}
i_{\phi} i_{\phi} C^{(p)}=\frac{1}{2}[\phi^i, \phi^j] C^{(p)}_{ji}
\end{equation}
and so the Chern-Simons action can be written schematically in the form
\begin{equation}
S_{CS} \sim \int d^{p+1} \xi  \left(C^{(p+1)} + i \lambda i_{\phi} i_{\phi} C^{(p+3)}+\ldots \right).
\end{equation}
Our interest is in coupling this system to four-dimensional Einstein gravity. Since in our solution the dilaton is fixed at zero - there is no concern about 
string frame effects\footnote{Although this means that the string coupling is essentially unity.}. 
The transverse scalars in this instance are now matrices, and we choose the fields to be proportional to generators of a non-Abelian gauge group. Since there
is often a transverse $S^2$ present in these models, we choose the group to be $SO(3) \sim SU(2)$ to reflect this transverse symmetry. As a result we
have the following ansatz for our fields
\begin{equation}
\phi^i = R(t) \alpha^i
\end{equation}
where $\alpha^i$ are the irreducible representation of the $SU(2)$ algebra. We then plug this ansatz into the action and follow the prescription discussed in \cite{ward} 
to obtain the action.
Let us simply state the relevant results. The diagonal components of the energy momentum tensor can be written as follows
\begin{eqnarray}
\rho &=& N T_3 \left(W h^4 \gamma - h^4 + V(\phi) \right) \nonumber \\
P &=& -N T_3 \left(Wh^4 \gamma^{-1} - h^4 + V(\phi) \right)
\end{eqnarray}
where $\gamma = (1-\dot{\phi}^2 / (h^4 T_3))^{-1/2}$ and $W = (1+4 \phi^4 /(h^4 \lambda^2 T_3^2 C_2))^{1/2}$ are the relativistic factor and the 
fuzzy potential terms respectively. Note that $C_2$ is the quadratic Casimir of $SU(2)$, which is related to the number of branes through the relation
$C_2 = N^2 -1$. We will keep the explicit dependence on the Casimir, although it should be noted that we have already assumed that $1/N^2$ terms
are negligible in obtaining the above expressions.
The scalar potential $V(\phi)$ has been included in order to account for other brane/flux interactions which may be present. 
The expression for the sound speed is also the same as in single brane models \cite{dbiinflation}
\begin{equation}
C_s^2 = \frac{1}{\gamma^2}
\end{equation}
which implies that the level of non-Gaussian fluctuations should be independent
of $N$, which is a somewhat surprising result\footnote{The $N$ dependence plays a role in setting the scale of the amplitude fluctuations \cite{ward}.}.
The fact that the large $N$ limit gives the same speed of sound as the single brane model is actually not surprising, since (as we will show in the next section)
this configuration actually has a dual interpretation in terms of a single wrapped brane \cite{myers}. 
As is well known, in the full warped deformed conifold solution \cite{klebanovstrassler} the second Betti number of this background is zero which means that there are no stable non-trivial two-cycles within the geometry. This implies that any brane wrapped along this cycle can shrink to a point. However one way to stabilise the brane on this cycle is to turn on $F1$-string charge. Thus our string solutions will typically have to carry some
extra $U(1)$ gauge theory on their world volumes.

The important prediction of DBI inflation is that it can lead to large levels of non-Gaussian fluctuations \cite{dbiinflation, constraints, related}. The current sensitivity of the
WMAP data at the $0.95$ confidence level\cite{data} only places the minimal bound on these fluctuations to be 
\begin{equation}\label{eq:ngbound}
-256 < f_{nl} < 332. 
\end{equation}
If one also assumes that the scalar perturbations are given by $\zeta = \zeta_L (1-3/5 f_{nl} \zeta_L)$, where $\zeta_L$ denotes the linearised Gaussian perturbations, then
one can derive the following three-point function in momentum-space \cite{nongaussianities}
\begin{equation}
<\zeta({\bf k_1}) \zeta({\bf k_2}) \zeta({ \bf k_3})> = (2 \pi)^7 \delta^3({\bf k_1}+{\bf k_2}+{\bf k_3})\frac{\sum_i k_i^3}{\Pi_i k_i^3}\left(-\frac{3f_{nl}}{10}(P^{\zeta}_k)^2 \right)
\end{equation}
where $P^{\zeta}_k$ is the scalar amplitude in momentum space. Now the non-Gaussianity amplitude $f_{nl}$ has six contributions, of which only two are relevant
for DBI inflation since the others are of order of the slow roll parameters. In the equilateral triangle limit, where all three momenta are equal, we can
approximate the amplitude of these fluctuations through the following expression
\begin{equation}\label{eq:nongaussian}
f_{nl} = \frac{35}{108}\left(\frac{1}{C_s^2}-1 \right)-\frac{5}{81}\left(\frac{1}{C_s^2}-1-2\Lambda \right)
\end{equation}
where the $\Lambda$ function is determined through the following relation
\begin{equation}
\Lambda = \frac{X^2 P_{,XX}+\frac{2}{3}X^3 P_{,XXX}}{XP_{,X}+2X^2P_{,XX}}
\end{equation}
where we have defined $X=\dot{\phi}^2/2$. For slow roll inflation, the sound speed is always unity and moreover there are no contributions to the $\Lambda$ term
therefore $f_{nl} \sim 0$. However for DBI inflation, both for a single brane and for a large number of coincident branes, the sound speed is small. 
The result\footnote{We are using the conventions of \cite{nongaussianities}, which maybe of the opposite sign to those employed by the WMAP normalisation \cite{data}.} is that $f_{nl} \sim 0.32 \gamma^2$ and therefore could be observable for large $\gamma$. In practice
this provides us with a tight constraint on the allowed range of $\gamma$, which we can use to tune the inflationary scale. One important thing to note about
this result is that the solution is independent on the warp factor of the background, and is therefore a universal result even though the inflationary solutions are 
background dependent. Of course $\gamma$ is itself a function of the warping, however once we treat this as being a variable in itself we see
that there is no additional warp factor dependence.
However the running of the non-Gaussian amplitude, which is approximately given by $n_{nl}-1 \sim -2s$ where $s = \dot{C_s}/(C_s H)$, \emph{is} sensitive
to the particular choice of background.

The relevant cosmologically observable scales are set through the size of the Hubble parameter, which we define as
\begin{equation}\label{eq:friedmann}
H^2 = \frac{\rho}{3 M_p^2}
\end{equation}
where we are using the reduced Planck mass as is usual in String cosmology. An important relationship between the four-dimensional physics and the
ten-dimensional physics is set through the definition of the four-dimensional Planck scale \footnote{See \cite{kallosh} for recent discussion of this point.}
\begin{equation}
M_p^2 = \frac{V^{w}_6}{\kappa_{10}^2}
\end{equation}
where $V^w_{6}= \int d^6 \chi \sqrt{g} h(\chi)$ is the warped six-dimensional volume, and $\kappa_{10}^2 =\frac{1}{2} (2\pi)^7 g_s^2 \alpha'^4$ is the ten-dimensional
Newtons constant. This relationship is crucial when discussing the Lyth bound on the allowed range for the inflaton field. In the single
brane models it was shown that a relativistic inflaton was not compatible with this bound, when normalised to the WMAP 3 data. However the multi-brane
model contributes an additional factor of $\sqrt{N}$ to the allowed field range, and is therefore able to by-pass these stringent conditions and therefore
still provides a testable prediction \cite{ward, constraints, wrappedbranes}. 
Of course the fact that large $N$ will also lead to back-reaction on the geometry implies that our solution must
be extremely fine-tuned, But we will return to this issue in a later section.
As an approximation we can assume that the warped volume factorises into the bulk (CY) contribution, and the throat contribution. It is readily noted that the
throat volume is given by 
\begin{equation}
\tilde V_6 \sim \rm{Vol(X_5)} \int_0^{\rho_c} d\rho \frac{\rho^5}{h^4}
\end{equation}
where $\rho_c$ corresponds to the UV cutoff in the throat. In some models this cutoff will be taken to the the place where the throat is glued to the 
CY, whereas in other models the cutoff will represent the limit of reliability of the theory. In the expression note that $\rm{Vol(X_5)}$ is dependent
on the explicitly choice of five-dimensional manifold, but its volume will always scale like $a \pi^3$ where $a$ is some constant which is in the range 
$\mathcal{O}(10^0-10^1)$.
As a result the Planck mass can be assumed to be bounded through the relationship
\begin{equation}
M_p^2 > \frac{\tilde V_6}{\kappa_{10}^2}
\end{equation}
For inflationary trajectories we typically demand that $N T_3 V$ dominates the contribution to the Hubble scale in (\ref{eq:friedmann}). In the usual DBI model this implies that $V >> h^4(\gamma-1)$, which can be achieved even for relativistic rolling provided that the warp factor suppression is large enough. However one must be more careful when in the slow roll regime if this is not satisfied. Generally the warping will be 
exponentially suppressed, and is a function of the flux ratio. With appropriate fine tuning of these fluxes one would anticipate
that the warping can be sufficiently small.
In our model we find a slightly modified bound compared to the previous case given by $V >> h^4(W\gamma-1)$, and we must also ensure that $M/N >>1$ in order for
the back reaction to be negligible.
Now the fuzzy potential $W$ is bounded by unity from below, and is typically an increasing function of $\phi$ 
(depending upon the interplay with the warp factor). This means that the relevant scale is now set by $W\gamma$
and not just $\gamma$. This gives rise to a two-dimensional parameter space, and therefore access to a larger range of
inflationary trajectories.

First notice that for solutions where $W \sim 1$, the potential constraint reduces to the usual $D3$-brane models. However we must
also supplement this with the W-condition which imposes a bound on the inflaton range
\begin{equation}
V(\phi) >> h^4(\gamma-1) \hspace{0.5cm} \to \hspace{0.5cm} \phi^2 << \frac{M_s^2 h^2 \sqrt{C_2}}{8\pi^2 g_s}.
\end{equation}
This illustrates the sensitivity of the field to the warping, the string scale and the coupling constant. For solutions where the warp factor
approaches a constant (i.e position independent) the solution will be sensitive to the UV cutoff. Let us assume that the maximal allowed
value for the inflaton is given by $\phi_c = \rho_c \sqrt{T_3}$. In turn this means that the W-bound becomes a bound on the number of branes, and
we see that
\begin{equation}
N_{\rm const}  < \frac{M_s^2 \rho_c^2}{\pi h^2}
\end{equation}
suggesting that it is more preferable for the warp factor to be constant over longer distances. For semi-explicit string models the
warping will typically be of the form $h \sim h_0 \pm h_1 \rho^{\alpha} + \ldots$, in which case the cutoff corresponds to the 
maximal allowed value of $\rho$ that allows us to neglect the $\rho^{\alpha}$ terms.
Alternatively we can consider backgrounds such as $AdS_5 \times X_5$, in which case the normalised warping is given by $h \sim \phi/(R \sqrt{T_3})$ 
where $R^4 = 4\pi g_s M l_s^4$ is the usual curvature of the $AdS$ geometry
and $M$ is the total background flux.
Combining the W-condition with the flux constraint gives us the following (weak) bound on $N$, namely
\begin{equation}
N_{AdS_5} >> \frac{4 g_s}{\pi}.
\end{equation}
without having to resort to imposing the Lyth bound \cite{constraints}.

The $W \sim 1$ limit essentially maps onto the single-brane case, therefore the more interesting limit is to consider solutions where $W >> 1$, 
implying that the bound on the potential becomes:
\begin{equation}
V(\phi) >> \frac{8 \pi^2 h^2 \phi^2 g_s}{M_s^2 \sqrt{C_2}}
\end{equation}
where we have used the fact that the W-condition demands that $\phi^2 >> M_s^2 h^2 \sqrt{C_2}/(8 \pi^2 g_s)$.
This latter solution requires the warp factor to be extremely small if we wish to consider IR inflation \cite{irinflation}, as we still want the solution to
consist of perturbative string states. 

On cosmological scales we see from (\ref{eq:friedmann}) that inflation will impose an additional bound on the number of branes, since we require
$H^2 >> m_{\phi}^2$, where $m_{\phi}$ is the inflaton mass which arises from the subleading terms in the potential (at least for IR inflation). This 
means we can write a bound on $N$ through the relation
\begin{equation}
N >> 24 \pi^3 g_s \left(\frac{ M_p^2 m_{\phi}^2}{M_s^4}\right).
\end{equation}
Note that this is sensitive to the splitting between the inflationary scale and the string scale. For us to be confident about neglecting the
backreaction we require $N$ to be as small as possible whilst still allowing the $1/N^2$ terms to be negligible. This suggests that the string
scale should be high in these models, in which case the inflaton mass only need be of order the GUT scale. In a fully compactified theory,
we could also use the F-theory tadpole constraint to view this as an additional constraint upon the ratio $m_{\phi}^2/M_s^4$, since the fluxes
are bounded by the Euler number of the particular Calabi-Yau \cite{geometricflux, constraints}.

Let us restrict ourselves to the case of relativistic motion, where we approximate $\dot \phi \sim h^4 T_3$ which gives us
the inflaton equation of motion for a choice of warp factor. For constant warping the solution is $\phi(t) \sim \phi_0 + h^2 \sqrt{T_3} (t-t_0)$, whilst
for $AdS$ space it becomes $\phi \sim \phi_0 \rm{exp}(\sqrt{T_3}(t-t_0)/R)$.
The relativistic limit is of interest because this is the regime where the 
non-linearities play an important role. For the non-relativistic case we refer the reader to \cite{slowroll}.
The equation for the conservation of energy gives us a term on the RHS which goes like $-3 H NT_3 h^4 W \gamma (1-\mathcal{O}(\gamma^{-2}))$
however we can neglect the $1/\gamma^2$ terms as we are assuming the relativistic limit. Combining this with the Hamilton Jacobi formalism, where
we assume a monotonic trajectory for the inflaton field, we see that the first cosmologically relevant  parameter becomes \cite{perturbations}
\begin{equation}
\epsilon_1 = -\frac{\dot{H}}{H^2} \sim \frac{2 M_p^2}{N W_* \gamma_*} \left(\frac{H'_*}{H_*} \right)^2
\end{equation}
which is a slight modification of the usual DBI 'fast roll' parameter. In the solution above note that a prime denotes
differentiation with respect to $\phi$, and $*$ denotes that the parameter is evaluated at horizon crossing. 
The other two relevant terms are written below
\begin{equation}
\epsilon_2 = \frac{\ddot{\phi}}{H \dot{\phi}} \hspace{1cm} \epsilon_3 = \frac{\dot{F}}{2HF}
\end{equation}
where $F=P_{,X} + XP_{,XX}$ as usual. The resulting expressions reduce to the following
\begin{eqnarray}
\epsilon_2 &=& -\frac{2 M_p^2}{N W_* \gamma_*}\left(\frac{H'_*}{H_*} \right)\left(\frac{H''_*}{H'_*}-\frac{W'_*}{W_*}-\frac{\gamma'_*}{\gamma_*} \right) \nonumber\\
\epsilon_3 &=& -\frac{M_p^2}{NW_* \gamma_*} \left(\frac{H'_*}{H_*} \right) \left(\frac{W'_*}{W_*}+\frac{3\gamma'_*}{\gamma_*} \right).
\end{eqnarray}
Note that typically we will find $W'/W \ge 0$ which therefore makes the relevant parameters more negative.
Assuming the validity of the fast roll expansion, namely that $\epsilon_i << 1$, we see that the spectral indices for the
curvature and tensor perturbations may be written as follows
\begin{eqnarray}
n_s &=& 1 - 2 (2 \epsilon_1 + \epsilon_2 + \epsilon_3) \nonumber \\
n_t &=& -2 \epsilon_1
\end{eqnarray}
or in terms of the generalised background parameters 
\begin{eqnarray}
n_s &=& 1- \frac{2M_p^2}{N W_* \gamma} \left(\frac{H'_*}{H_*} \right) \left(\frac{4 H'_*}{H_*}- \frac{H''_*}{H'_*}- \frac{2\gamma'_*}{\gamma_*} \right) \nonumber \\
n_t &=& -\frac{4M_p^2}{NW_* \gamma_*} \left(\frac{H'_*}{H_*} \right)^2.
\end{eqnarray}
Note that because the solutions of interest correspond to large $\gamma$, we see that $n_t$ is actually independent of the fuzzy potential.
One can see this because once we write $\gamma$ as a function of $\phi$ we find that in the ultra-relativistic regime
\begin{equation}
W\gamma \sim \frac{2 M_p^2 |H'|}{\sqrt{T_3} h^2 N}
\end{equation}
and therefore the relevant scalar tilt is only a function of $H$ and its derivatives.
Now the relevant amplitudes for these perturbations in a de-Sitter background have been calculated for all the most general
cases of interest \cite{perturbations}. We repeat them here for convenience
\begin{eqnarray}
\mathcal{P}_s^2 &\sim& \frac{H^2_*}{8\pi^2 M_p^2 \epsilon_{1*} C_{s*}} \nonumber \\
\mathcal{P}_t^2 &\sim& \frac{2H^2_*}{\pi^2 M_p^2}.
\end{eqnarray}
Note that, as usual, the inflaton doesn't mix with gravitational modes and so there is no additional field dependence in
this amplitude besides the contribution to the Hubble scale.

Typically in IR models of DBI inflation \cite{irinflation, constraints}, the inflaton potential will be of the form $V \sim V_0 - \tilde m^2 \phi^2 + \ldots$, where we have
omitted the subleading corrections. In the notation of this paper $\tilde m$ has units of $(\rm{mass})^{-1}$ and therefore corresponds to some
length scale, which is different to the inflaton mass scale \footnote{We hope this will not further confuse the reader. We have tried to keep the overall dimensionful quantities as pre-factors throughout the paper.} $m_{\phi} = \sqrt{T_3} \tilde m$. In any event this leads to the approximation
\begin{equation}
\left(\frac{H'_*}{H_*} \right) \sim -\frac{\tilde m^2 \phi_*}{V_0}.
\end{equation}
Using this as our basis we can work out the detailed inflationary dynamics of this configuration in arbitrary backgrounds once we specify the
form of the harmonic function. An interesting example is when we consider $h \sim $constant as in the Klebanov-Strassler (KS) throat \cite{klebanovstrassler}. 
In terms of observable signatures, the most 
useful turns out to be the tensor index, which is given by
\begin{equation}
n_t \sim -2 N V_0^2 \left(1 - \frac{3 M_p^2 \tilde m^2 N_e}{N^2 V_0^2}\right) 
\end{equation}
where $N_e$ is the number of e-foldings before the end of inflation. If the term in brackets is close to zero, then $\epsilon_1 \sim 0$ and 
inflation occurs rapidly. However if this term is still appreciable then we see that the tensor index goes like $N$ and is therefore
an interesting observable. This has been discussed at length elsewhere \cite{ward}, so we will not mention it further here.
\subsection{The dual picture}
Let us now demonstrate how this configuration is related to that of a wrapped $D5$-brane \cite{wrappedbranes}. 
In order for this configuration to exist we must ensure that there
is two-cycle within the transverse space that our $D5$-brane can be wrapped upon. If we factorise the compact metric into products of spheres, then the metric can
be factorised into $d\psi^2 + \sin^2 \psi d\theta^2 + \ldots$ and we will choose our internal embedding coordinates to be $(\psi, \theta)$. The remaining
world-volume coordinates are extended in the non-compact directions as usual. For this wrapped configuration to be dual to the one introduced in the previous
section, we must ensure that there is a non-zero $U(1)$ 'magnetic' flux on the world-volume. In order not to break Lorentz invariance, this flux must
lie along the compact directions. In fact the introduction of the magnetic charge will not stabilise the configuration, as it is well known that non-zero
electric charge is required for full stabilisation of the cycle.
For simplicity we shall also only consider the leading order Chern-Simons contribution to the action. The introduction of world-volume flux also allows
for non-trivial contributions to the pullback of the $C^{(4)}$, but we will ignore these effects in this section.
Calculation of the DBI part of the action results in the following expression
\begin{equation}
S = -T_5 \int d^6 \xi h^4 \sqrt{1-h^{-4} \dot{\rho}^2}\sqrt{h^{-4}\rho^4 \sin^2 \psi + \lambda^2 F_{\psi \theta}^2}.
\end{equation}
Since the magnetic field is fully localised in the compact directions, it should be proportional to the cycle volume so we will take the following
ansatz for the flux where there are $N$ units of charge
\begin{equation}
F^{(2)} = N \omega_2
\end{equation}
where $\omega_2$ is the two form on the transverse $S^2$.
This choice of field simplifies the full action tremendously and we can find
\begin{equation}
S = -2 \pi T_5 \int d^4 \xi \left(h^4 \sqrt{1-h^{-4} \dot{\rho}^2}\sqrt{h^{-4}\rho^4 + \lambda^2 N^2} - h^4 \lambda N \right)
\end{equation}
where we have included the contribution coming from the Chern-Simons term, and also integrated out the compact directions.
If we now factorise this expression and use the following relation $T_5/T_3 = M_s^2/(4 \pi^2)$, and also switch to the canonical field
description where $\phi = \rho\sqrt{T_3}$, then the brane contribution to the Lagrangian density becomes
\begin{equation}
\mathcal{L} = -NT_3 h^4\left(\gamma^{-1} W -1 \right)
\end{equation}
where both $W$ and $\gamma$ are the functions derived in the previous section (provided one takes the large $N$ limit). 
Another consequence of the nature of this picture is that we can understand why the backreaction of the $D5$-brane is non-negligible, since
the dual picture consists of $N$-coincident branes - which perturb the background due to their cumulative mass.
One should also note that this configuration should also be dual to a $D7$-brane wrapping a non-trivial four-cycle, provided that the $D7$-brane has
a non-vanishing second Chern-Class so that the action will contain a coupling of the form $\int C^{(4)} \wedge F \wedge F$. Therefore one could start
building inflationary models from the $D7$-brane perspective by including additional world-volume flux in the compact directions.

The dual nature of this model suggests that we can consider the cosmology of either $N$ coincident $D3$-branes, 
or a single wrapped $D5$-brane with $N$ units of magnetic charge. This latter description
has recently been investigated in \cite{wrappedbranes}. One important consequence of this description is that the moving $D5$-brane naturally
excites its $U(1)$ world-volume gauge fields. This suggests that the multi-brane configuration will also excite its world-volume fields, which will now of course
be charged under the $U(N)$, and therefore the standard model degrees of freedom will be reheated. In the single brane scenario, the inflaton sector
is gravitationally coupled to the standard model sector and the expectation is that reheating will occur through tunneling of the KK modes between the inflaton
throat and the standard model throat. Whilst this is a reasonably robust mechanism, the $U(1)$ gauge boson associated with the open string modes on the 
$D3$-brane will remain a relevant degree of freedom and therefore at least some of the inflationary energy will go into exciting these string states \cite{reheating}.
Whilst this remains a hidden sector from the standard model perspective, the open string excitations should still provide a definite physical signature and
it would be interesting to work this out in detail.
The multi-brane model does not suffer from this problem, since the inflationary energy is expected to be dumped into open string states at the end of inflation with
only a small amount emitted via tunneling. However the brane configuration will be near the $UV$ end of the throat and so one must understand how
the branes will backreact on this internal geometry in order to discuss their evolution in this region. Moreover since the branes are all assumed to be parallel,
there will not be any chiral fermions in the spectrum. There are potentially two ways in which this can be alleviated. Firstly one could assume that
the dynamical brane stack intersects with another (stationary) stack localised at the tip of another throat, in which case the symmetry group will be enhanced 
to $U(N)\times U(N')$ which will give rise to both adjoint and fundamental matter. The second possibility is that although the branes are within a string
length of one another, they may not be exactly parallel and therefore could be sensitive to tidal forces or the exact profile of the unstable potential. 
This means that some of the branes may be intersecting, but at initially unobservable scales. Alternatively one may imagine that the dynamically induced fluctuations
will lead to some branes intersecting. It is important to develop these ideas in more detail in order to understand how the inflaton sector couples to the standard model,
since this is a particularly weak area for these models \cite{reheating}.

Since we know from the wrapped $D5$-brane picture that backreactive effects can be important, we should also try to understand how they might emerge in the
multi-brane case which we attempt to address in the following section.
\section{Including $1/N$ Corrections}
Our results have been written explicitly in terms of $N$ and $C_2 = N^2-1$, and so we could clearly incorporate $1/N$ corrections
simply by keeping the $1/N^2$ pieces of the quadratic Casimir. However we must seek to ensure that there are no other corrections
appearing at this order which could cancel these terms. We can do this by considering the corrections coming from the symmetrised trace prescription.

Whilst the full non-Abelian DBI action remains unknown, we know that the Myers prescription agrees with the wrapped $D5$-brane description in
the large $N$ limit \cite{myers}. We also know from string scattering amplitudes that we must include some symmetrisation \cite{tseytlin} if we are to project 
out unwanted terms at leading order. 
This suggests that we should at least consider the possibility that symmetrisation terms could play a role in
the full description of the non-Abelian action. This our proposal here is certainly not the full solution, but will comprise part of it.

Recall that our main focus was on the large $N$ limit in obtaining the action for the inflaton. 
We would like to go beyond this approximation to capture the next leading order corrections \cite{finiten}, and see if they alter the dynamics of the solution. 
The question we need to address is therefore how does symmetrisation affect the gauge trace?
Since our $D3$-brane solution has no world-volume gauge fields, the only terms to be traced over are the generators of the $SO(3)$ algebra. Therefore our
question reduces to a simpler one, namely calculating the symmetrised trace over these generators. This is actually just a question
of combinatorics, however we can use an alternate description in terms of chord diagrams, or so called "bird track" diagrams.

Let us introduce the following graphical description of the group generator $\alpha^i$ using 
\begin{displaymath}\label{ch1}
{(\alpha^i)}^a_b \quad = \quad
\begin{picture}(2,2)
\put(1,0){\line(0,1){1.5}}
\put(0,-0.5){\makebox(0,0){${}^a$}}
\put(2,-0.5){\makebox(0,0){${}^b$}}
\put(1,1.7){\makebox(0,0){${}^i$}}
\thicklines
\put(0,0){\line(1,0){2}}
\end{picture}
\end{displaymath}
where the $i$ runs from $1 \to 3$ and $a, b$ are matrix labels. to reflect the fact that we are in the adjoint representation.
Now our generators will come in pairs, so our problem amounts to determining the solution of $STr(\alpha^i \alpha^i)^n$ in order to
calculate the full solution once we expand the square root terms of the DBI action.
Let us focus on the case $n=1$ initially. In this case we must join multiply the two generators together, and then trace over the
gauge index. In our graphical notation this amounts to joining the free ends of the line to make a circle. Thus we see that
\begin{equation}\label{traceofn1}
\frac{1}{N} Tr(\alpha^i \alpha^i) =
\ssk \Picture{\FullCircle\EndChord[3,9]} = \hspace{0.1cm} C_2
\end{equation}
\\
\noindent which is the only possible diagram that we can form. Note that we have pulled out the factor of $N$ coming from the trace
over the identity matrix, which is a standard convention employed in the literature.
Now let us consider the case where $n=2$, which will have two different diagrams as follows 
\begin{equation}\label{ch2} \abb
\Tr(\alpha^i\alpha^i\alpha^j\alpha^j)
=\ssk \Picture{\FullCircle\EndChord[3,6]\EndChord[0,9]} \qquad
\Tr(\alpha^i\alpha^j\alpha^i\alpha^j) =\ssk
\Picture{\FullCircle\EndChord[0,6]\EndChord[3,9]}.
\end{equation}
\\
However if we also keep track of the relative weighting of each diagram we see that the first contributes a weighing of $2/3$, whilst the
second is $1/3$. It is these weighting factors which are important for the symmetrisation procedure. At this stage we want to turn
our diagrams above into something algebraic, since we know that they should correspond to some function of the Casimir. In fact the
first diagram is simply the direct product of two copies of (\ref{traceofn1}), and so this diagram is equal to $2 C_2^2/3$ when we include
the weighting factor. The second diagram is more complicated, however we can remove one of the internal lines using
\begin{equation}
\Picture{\FullCircle\EndChord[0,6]\EndChord[3,9]} = \ssk  (C_2 - 4) \hspace{0.2cm} \Picture{\FullCircle\EndChord[3,9]}
\end{equation}
 
\noindent
which reduces the diagram to $C_2(C_2 - 4)$. If we add the contribution coming from both terms then we find that
\begin{equation}
\frac{1}{N} STr(\alpha^i \alpha^i)^2 = C_2^2 \left(1-\frac{4}{3 C_2} \right).
\end{equation}
What about the next order terms coming in at $n=3$? There are five unique diagrams contributing at this order, which are shown below
\begin{equation}
\ssk
\Picture{\FullCircle\EndChord[1,11]\EndChord[3,9]\EndChord[5,7]}\ssk,\quad
\Picture{\FullCircle\EndChord[2,8]\EndChord[4,10]\EndChord[0,6]}\ssk,\quad
\Picture{\FullCircle\EndChord[2,6]\EndChord[4,10]\EndChord[0,8]}\ssk,\quad
\Picture{\FullCircle\EndChord[6,8]\EndChord[10,0]\EndChord[2,4]}\ssk,\quad
\Picture{\FullCircle\EndChord[4,8]\EndChord[3,11]\EndChord[9,1]} \nonumber
\ssk \\[20pt]
\end{equation}
\noindent
where we have omitted the weighting factors of each diagram. The decomposition occurs in much the same way as before and the final
result can be written as follows
\begin{equation}
\frac{1}{N} STr(\alpha^i \alpha^i)^3 = C_2^3 \left(1-\frac{4}{C_2}+\frac{16}{3C_2^2} \right).
\end{equation}
One immediate thing to note is that the leading term in these expansions goes like $C_2^n$, which is in fact the only
coefficient that gets picked out in the large $N$ limit. The other terms are clearly the sub-leading corrections we are looking for.
At level $n=4$, there are $18$ different chord diagrams to draw, which becomes $105$ diagrams at level five and so on. Each corresponding
value of $n$ contributes a larger set of diagrams, much in the same way as the Feynman expansion. 
Let us write down a series of definitions which help to simplify things:
\begin{itemize}
\item Let $I$ denote the number of intersections of a pair of chords.
\item Let $T$ denote the number of triple intersections of three chords.
\item Let $Q$ denote the number of quadratic intersections of four chords in the shape of a box.
\end{itemize}
Every chord diagram $D(n)$ can be written in terms of these intersection numbers as follows
\begin{equation}
D(n) = C_2^n - 2I C_2^{n-1} + 2 C_2^{n-2} \left( I(I-1) + 4Q - 2T\right) + \ldots
\end{equation}
where there are higher order terms which we are suppressing.
It can be shown that these terms can be summed to give the leading order terms for the symmetrised trace
\begin{equation}
\frac{1}{N} STr(\alpha^i \alpha^i)^n = C_2^n - \frac{2}{3}n(n-1) C_2^{n-1} + \frac{2}{45}n(n-1)(n-2)(7n-1) C_2^{n-2}+\ldots.
\end{equation}
If we regard this as a differential operator acting on some function of the Casimir such that
\begin{equation}
STr(\alpha^i \alpha^i)^n = D C_2^n
\end{equation}
then we can see that
\begin{equation}
STr F(\alpha^i \alpha^i) = \sum_{n=0}^{\infty} F_n D C_2^n = D F(C_2)
\end{equation}
where we write
\begin{equation}\label{eq:operator}
D := N \left(1-\frac{2C_2}{3} \frac{\partial^2}{\partial C_2^2} + \frac{8C_2}{9} \frac{\partial^3}{\partial C_2^3} + \ldots\right)
\end{equation}
where there are higher order terms that we have neglected. Therefore we see that the leading order term is just $N$ multiplied 
by the original function of the Casimir, which exactly corresponds to the large $N$ limit of the DBI action. The next to
leading order terms can be determined through the action of the differential operator.

Note that in the dual $D5$-brane picture we expect these corrections to correspond to non-commutative deformations of the gauge field
structure,  which in principle can be determined through the use of a star product on the world-volume.
In what follows we will restrict our analysis to solutions where the warp factor becomes constant in the IR, such as in the KS geometry \cite{related, klebanovstrassler}.
This is not the most general case one could consider, but the calculation of the corrections is hampered in this instance by the additional dependence of the
warp factor upon powers of the quadratic Casimir. 

Typically in IR models of DBI inflation, the last $60$ e-folds will be occur when the branes move away from the tip of the throat, and therefore the warp factor will
play an important role. For the UV scenarios - which we can also include in this analysis, the constant warping is required for the last 60-efolds of inflation
to occur. Let us assume a general form for the warp factor $h(\rho)$ and calculate the full corrections to the DBI Lagrangian including the $1/N$ terms.
It will be useful to define the following variables
\begin{eqnarray}
\alpha &=& 1-\frac{4 \delta h_c C_2}{h} \nonumber \\
\beta &=& 4C_2\left(\frac{\delta h_c^2}{h^2}-\frac{\delta h_{cc}}{h} \right)-\frac{4\delta h_c}{h}
\end{eqnarray}
where in this notation $\delta h_c$ corresponds to a derivative with respect to $C_2$. 
However since we know that $h=h(\rho)$, and that $\rho^2=\lambda^2 \hat{R}^2 C_2$ through the definition of the physical radius of the 
fuzzy sphere, we can write the derivatives of the warp factor
explicitly in terms of derivatives with respect to the inflaton field $\phi$ through the identification
\begin{equation}
\delta h_c = \frac{h' \phi}{2 C_2}.
\end{equation}
As an example let us consider the case where the metric is $AdS_5 \times X^5$, and therefore the warp factor can be written as follows
$h= \phi /(\sqrt{T_3} R)$, where $R$ is the usual $AdS$ scale. The above expression then reduces to $\delta h_c = h/(2 C_2)$ and is therefore
suppressed by a factor of $C_2$ with respect to the original function. 

Therefore we can write the expression for the energy density with the $1/N$ corrections as follows
\begin{equation}\label{eq:fullenergy}
\rho = NT_3 h^4 \left(W \gamma -1 +V h^{-4} - F_1(W, \gamma) \right) 
\end{equation}
where the function $F_1$ is defined below
\begin{eqnarray}
F_1(W, \gamma) &=& \frac{\gamma}{6 C_2} \left(\frac{W^2-1}{W}(2C_2 \beta - 2\alpha + \alpha^2\frac{(W^2+1)}{W^2})+\frac{2\alpha^2(W^2-1)(\gamma^2-1)}{W} \right) \nonumber \\
&+& \frac{\gamma}{6C_2}\left(W(\gamma^2-1)[\alpha^2(3\gamma^2-1)-2\alpha-2\beta C_2]\right) \nonumber \\
&+& \frac{\gamma}{6 C_2 }\left(\frac{4C_2^2}{\gamma h^4}[\delta V'' -\frac{8 \delta V'\delta h_c}{h} - \frac{4 V \delta h_{cc}}{h} + \frac{20 V h_c^{2}}{h^2}]\right) \nonumber \\
&+& \frac{8h' \gamma}{3 h}\left(\frac{\alpha(W^2-1)}{W}+W\alpha(\gamma^2-1)+\frac{2C_2}{\gamma h^4}[\delta V' - \frac{4V \delta h_c}{h}] \right) \nonumber \\
&+& \frac{8 C_2}{3} \left(W\gamma-1+\frac{V}{h^4} \right)\left(\frac{3h_c^2}{h^2}+\frac{\delta h_{cc}}{h} \right)
\end{eqnarray}
Note that $\delta V', \delta V''$ corresponds to taking derivatives with respect to $C_2$ here, and not the inflaton field.
The corresponding solution for the pressure may be written as
\begin{equation}\label{eq:fullpressure}
P = -N T_3 h^4 \left(\frac{W}{\gamma}-1+\frac{V}{h^4}-F_2(W, \gamma) \right)
\end{equation}
where the correction function $F_2$ is defined to be as follows
\begin{eqnarray}
F_2(W, \gamma) &=& \frac{1}{6 \gamma C_2} \left(\frac{W^2-1}{W}(2C_2 \beta - 2\alpha + \alpha^2\frac{(W^2+1)}{W^2})-\frac{2\alpha^2(W^2-1)(\gamma^2-1)}{W} \right) \nonumber \\
&+& \frac{1}{6 \gamma C_2}\left(-W(\gamma^2-1)[\alpha^2(3\gamma^2-1)-2\alpha-2\beta C_2]+2W\alpha^2(\gamma^2-1)^2\right) \nonumber \\
&+& \frac{1}{6 \gamma C_2 }\left(\frac{4\gamma C_2^2}{h^4}[\delta V'' -\frac{8 \delta V'\delta h_c}{h} - \frac{4 V \delta h_{cc}}{h} + \frac{20 V \delta h_c^{2}}{h^2}]\right) \nonumber \\
&+& \frac{8\delta h_c }{3 \gamma h}\left(\frac{\alpha(W^2-1)}{W}-W\alpha(\gamma^2-1)+\frac{2C_2\gamma}{h^4}[\delta V' - \frac{4V \delta h_c}{h}] \right) \nonumber \\
&+& \frac{8 C_2}{3} \left(\frac{W}{\gamma}-1+\frac{V}{h^4} \right)\left(\frac{3\delta h_c^2}{h^2}+\frac{\delta h_{cc}}{h} \right)
\end{eqnarray}
These expressions clearly show the sensitivity of the solution to the warp factor, and therefore we should restrict ourselves to specific backgrounds
in order to understand how the corrections alter the physics.
\subsection{The limit of constant warping.}
The above expressions will clearly simplify when we assume constant warping, as in the Klebanov-Strassler geometry. After careful
computation, the respective energy and pressure densities including the $1/N$ corrections can be written using 
parameterisation invariant functions as follows
\begin{eqnarray}
\rho &=& NT_3 h^4 \left(W\gamma-1+\frac{V}{h^4}-\frac{\gamma F_1(W,\gamma)}{6C_2} \right) \nonumber \\
P&=& -NT_3 h^4\left(\frac{W}{\gamma}-1+\frac{V}{h^4}-\frac{F_2(W,\gamma)}{6\gamma C_2} \right)
\end{eqnarray}
where we have defined
\begin{eqnarray}
F_1(W, \gamma) &=& \frac{2(W^2-1)(\gamma^2-1)}{W} + 3W(\gamma^2-1)^2-\frac{(W^2-1)^2}{W^3} + \frac{4C_s^2 \delta V''}{\gamma h^4} \nonumber\\
F_2(W, \gamma) &=& \frac{(W^2-1)^2}{W^3}+\frac{2(W^2-1)(\gamma^2-1)}{W}+W(\gamma^2-1)^2-\frac{4\gamma C_2^2 \delta V''}{h^4} \nonumber
\end{eqnarray}
Clearly we see that the corrections are essentially all suppressed by powers of $1/C_2$ with respect to the leading order solution. 
The only place where one has to be careful is with the $\delta V''$ term, which in both cases is
enhanced by a factor of $C_2$. Of course for potentials which are essentially constant over the regime of interest, as assumed in IR inflation, 
these terms will vanish from the expressions above. 
Using these expression we can see that the inflationary constraint upon the potential dominance is now modified to read
\begin{equation}
V >> h^4 \left(\gamma \left(W - \frac{F_1(W, \gamma)}{6C_2}\right)-1 \right).
\end{equation}
If we set $W\sim 1$ in the above expressions then the constraint on $V$ in the large $\gamma$ limit is very weak due to the dependence
on $\gamma^4/C_2$.

We could enquire about how the Lyth bound is now altered by the presence of these $1/N$ corrections, however things rapidly become complicated. In the
notation of Lidsey and Huston \cite{constraints} 
we find that the correction term $P_3$ is not a function, but rather a \emph{functional} of both $P_1(\phi,X)$ and $P_2(\phi, X)$
where $X$ is the usual canonically normalised kinetic piece. As such one cannot easily extend their analysis to this more general case without first picking 
a restrictive gauge choice.
Since the parameter space of multi-brane inflation is larger than in the single brane case, we are able to find inflationary trajectories even when we include these
correction terms. What is more interesting from our perspective is to see how the $1/N$ terms alter the speed of sound and the non-Gaussian spectrum, since this
is where the signature of the model is important. Although the sound speed is not an observable quantity, it is an important parameter to 
calculate since fluctuations enter the horizon at $k C_s = a H$
The corresponding expression for the sound speed in a constantly warped background is found to be
\begin{equation}
C_s^2 = \frac{1}{\gamma^2} \frac{\left(W-\frac{1}{6C_2}\left\lbrack\frac{(W^2-1)}{W}\left\lbrace\frac{(W^2-1)}{W^2}-2(1+\gamma^2)\right\rbrace-W(\gamma^2-1)(1+3\gamma^2)\right\rbrack\right)}{\left(W-\frac{1}{6C_2}\left\lbrack\frac{(W^2-1)}{W}\left\lbrace2(3\gamma^2-1)-\frac{(W^2-1)}{W^2}\right\rbrace+3W(\gamma^2-1)(5\gamma^2-1)\right\rbrack \right)}
\end{equation}
which reduces to the usual solution $C_s^2 \sim 1/\gamma^2$ in the large $N$ limit. 
Let us investigate various limits of this expression in order to see if it imposes any conditional constraints upon the
dynamics. Firstly let us
consider the solution when $W \sim 1$, which would also be the case for a single $D3$-brane in the throat. At leading order in a large $\gamma$ expansion (assuming $\gamma^2>>1$)
we find that
\begin{equation}
C_s^2 \sim \frac{1}{\gamma^2} \left(\frac{2C_2 + \gamma^4}{ 2C_2 - 5 \gamma^4} \right) + \ldots
\end{equation}
where the ellipsis denote subleading terms. For this to be non-negative we require that the denominator satisfy a reality condition, which when combined
with the large $\gamma$ approximation implies that this expression is valid when $N^2>>7/2$, which is a rather weak bound on the number of $D3$-branes.
More interestingly we see that if we keep the Next to Leading Order (NLO) terms in the sound speed then we can solve the reality bound as a constraint on
$\gamma$ itself, which turns out to be
\begin{equation}\label{eq:gammaconstraint1}
\gamma^2_{*} < \frac{6}{15} \left(1 + \sqrt{1+\frac{15C_2}{6}} \right)
\end{equation}
which links $\gamma$ directly to the number of branes. This behaviour is a strictly subleading effect, and is not observed in the large $N$ (or single brane) case.
If we consider the non-relativistic expansion in this limit, then again we see that $C_s^2 \to 1$ as in ordinary slow-roll models - washing out the effect
of the $1/N$ corrections.

If we now consider the converse approximation, assuming slow roll from the start, then the functional form of the speed of sound appears to admit a non-trivial
solution which picks up corrections even in the 'squeezed limit' of zero velocity due to the non-trivial contribution from the fuzzy potential
\begin{equation}
C_s^2 \sim \frac{1+A}{1-A} \hspace{1cm} A = \frac{(1+3W^2)(W^2-1)}{6C_2 W^4}
\end{equation}
however one can check that this is an imaginary solution unless we also take the $W>>1$ limit.
Another interesting limit is the one capturing the non-Abelian structure of the theory, which assumes $W>>1$. In this case we find that all $W$ dependence
drops out of the sound speed leaving the following expression for all $\gamma$
\begin{equation}
C_s^2 \sim \frac{1}{\gamma^2}\left(\frac{1+ \gamma^4/(2C_2)}{1- \gamma^2(5 \gamma^2-4)/(2C_2)} \right).
\end{equation}
However this clearly imposes a bound on the physical values of $\gamma$, since this expression has a divergence at the critical limit where
$\gamma^2_{*} = 0.4 (1+ \sqrt{1+10 C_2/4})$, which is very similar to (\ref{eq:gammaconstraint1}) therefore when analysing this limit we must again 
ensure that $\gamma$ is below this bound in order for the
solution to be regarded as being physical. Of course
we clearly see that $\gamma^2_{*}$ increases as the number of branes increases, so we again find a non-trivial dependence of the relativistic factor on $N$.

One interesting observation is that the effect of the $1/N$ corrections acts to 'squeeze' the sound speed along the $\gamma$ direction. The function is
no longer monotonic in this limit, indeed we find that the function decreases with increasing $\gamma$ until it becomes small. However because
of the corrections the sound speed then increases to become large. Clearly this is not what is required for inflation. However note that when the
velocity is constant, $\gamma$ is also a constant which means that $W$ becomes important. Since $W$ effectively parameterises a flat direction of the
sound speed, we can still find inflation trajectories where the sound speed is small albeit for fixed $\gamma$. Once we move to larger $N$, the 
squeezing reduces and we find the sound speed is small over a larger range of $\gamma$ values.

Given the expression for the sound speed in these backgrounds, we can also calculate $f_{nl}$ - however this is a far more complicated function since the
additional corrections introduce new position/momentum interactions in the conjugate phase space. What we can easily observe about the form of $f_{nl}$ is its
behaviour as a function of $\gamma$, since it will be more sensitive to the new $\gamma$ interactions. Since both the sound speed and the 
non-Gaussianities can be calculated without imposing additional cosmological constraints, we will view $N$ as being a free parameter. Of course, as we saw in the
first section, smaller values of $N$ typically require $M_s \sim M_p$ and a low inflaton mass scale and therefore will correspond to finely tuned solutions. 
As one may expect the spectrum is sensitive to the precise value of $N$ and therefore we have plotted the non-Gaussianities for different numbers of branes as shown
in Figures [1-4], where we have assumed that the potential is roughly constant over the inflationary region.

For the first case with only ten branes, we are dropping terms of order $1/N^2$ which is at the one-percent level. This should be regarded as being the absolute limit
of our approximations. The backreaction will be reasonably under control in this instance provided that $M>>N$ is still satisfied. What it clear from the figures is that
there exists a turning point in the profile of $f_{nl}$ as a function of $\gamma$. Beyond the turning point, the spectrum becomes large and negative due
to the second term in (\ref{eq:nongaussian}) becoming dominant. This is also the region where the sound speed is starting to increase again, and will therefore
not necessarily allow for inflation. However near the turning point we know that the sound speed is small, and that inflation can occur along a trajectory
through ${W, \gamma}$ phase space, therefore the corrections predict a maximal value for $f_{nl}$ which is sensitive to the number of branes in the model.

As we increase $N$, the location of this maximum moves to larger values of $\gamma$ and the solution approaches the large $N$ behaviour. Again this is
because the sound speed is smaller over a larger range of $\gamma$. The location of this maximum is roughly at $\gamma \sim \sqrt{N/2}$, which is why
it is not visible in the large $N$ limit.
Note that the maximum amplitude is bounded from above due to the
competition between the two terms, and is also much smaller in amplitude than one may have anticipated. This is again a result of the
corrections, and it appears that larger $N$ leads to a larger observable signal.
Once we cross over a threshold number of branes, the turning point is pushed to larger and larger values of $\gamma$ and is therefore essentially unobservable.

The equation of motion for the inflaton can be determined using the Hamilton-Jacobi formalism, and the relevant energy-momentum tensor components. 
The Hubble equation correspondingly becomes
\begin{equation}
\dot{H}=-\frac{N T_3 h^4}{2 M_p^2\gamma}\left(W(\gamma^2-1)+\frac{1}{6C_2}\left(\frac{(W^2-1)^2(1+\gamma^2)}{W^3}-\frac{2(W^2-1)(1-\gamma^2)^2}{W}+W(1-3\gamma^2)(\gamma^2-1)^2 \right) \right) \nonumber
\end{equation}
where we have explicitly assumed that the $\delta V''$ term vanishes for simplicity.
Rather than solve the full equation of motion, let us consider a physical approximation where we assume $W\sim 1$ and that $\gamma$ is large. In this
instance we find the following expression at leading order
\begin{equation}
\dot \phi \sim -\frac{2 M_p^2 H'}{N \gamma} \frac{1}{(1+\gamma^4/(2C_2))}+\ldots
\end{equation}
where we have explicitly assumed that the field is a monotonic function. This expression is remarkably similar to the 
one derived in both the large $N$ and single brane cases \cite{dbiinflation, ward}. As a result the corresponding fast roll variable governing its dynamics is given by
\begin{equation}
\epsilon_1 \sim \frac{2M_p^2}{N \gamma}\frac{1}{(1+\gamma^4/(2C_2))} \left(\frac{H'}{H} \right)^2
\end{equation}
which is sensitive to the ratio $\gamma^4/(2C_2)$. Since we know that the $1/N$ corrections imply a non-trivial relationship between $\gamma$ and $N$ we 
will generally see that this correction term is typically small, although non vanishing, and will therefore act to suppress the slow roll parameter.
With appropriate tuning one can easily find inflation trajectories, as in the large $N$ limit \cite{ward}.
\begin{center}
\begin{figure}
\begin{minipage}[t]{.50\textwidth}
\includegraphics[width=\textwidth]{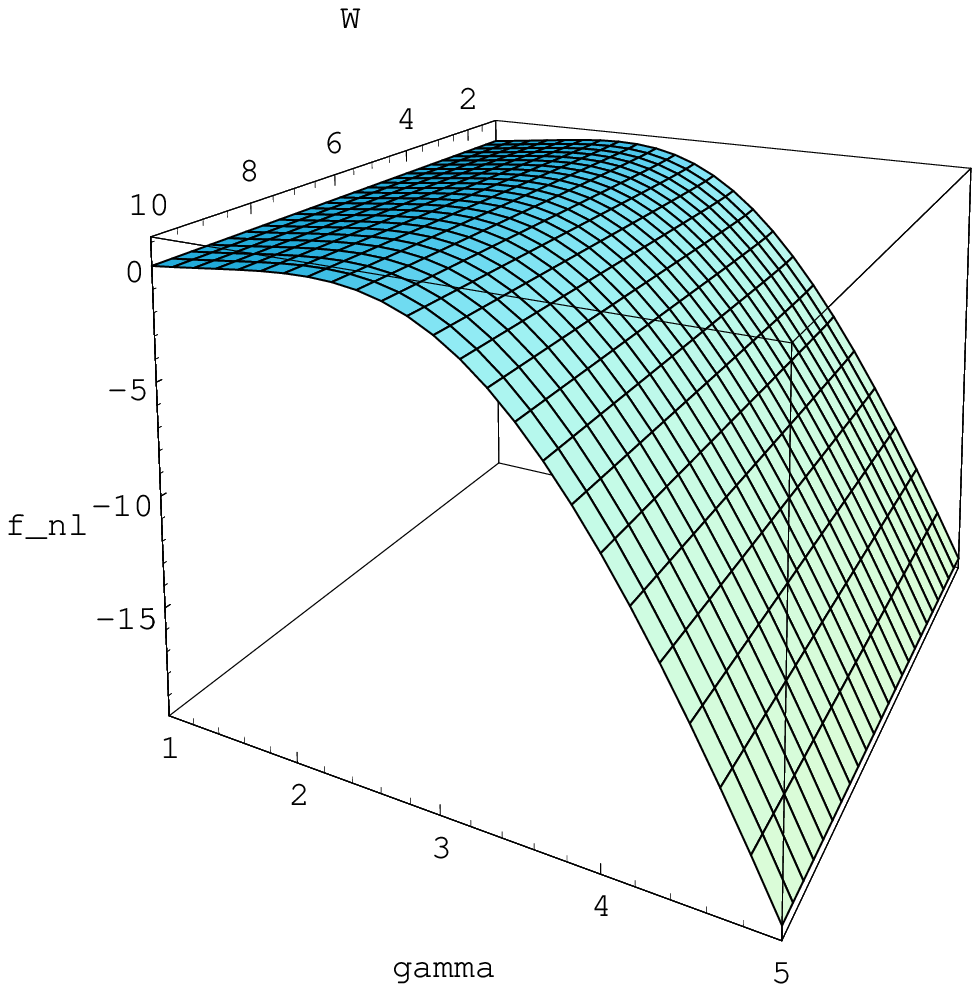}
\centering \texttt{Figure 1: $N=10$.}
\end{minipage}
\begin{minipage}[t]{.50\textwidth}
\includegraphics[width=\textwidth]{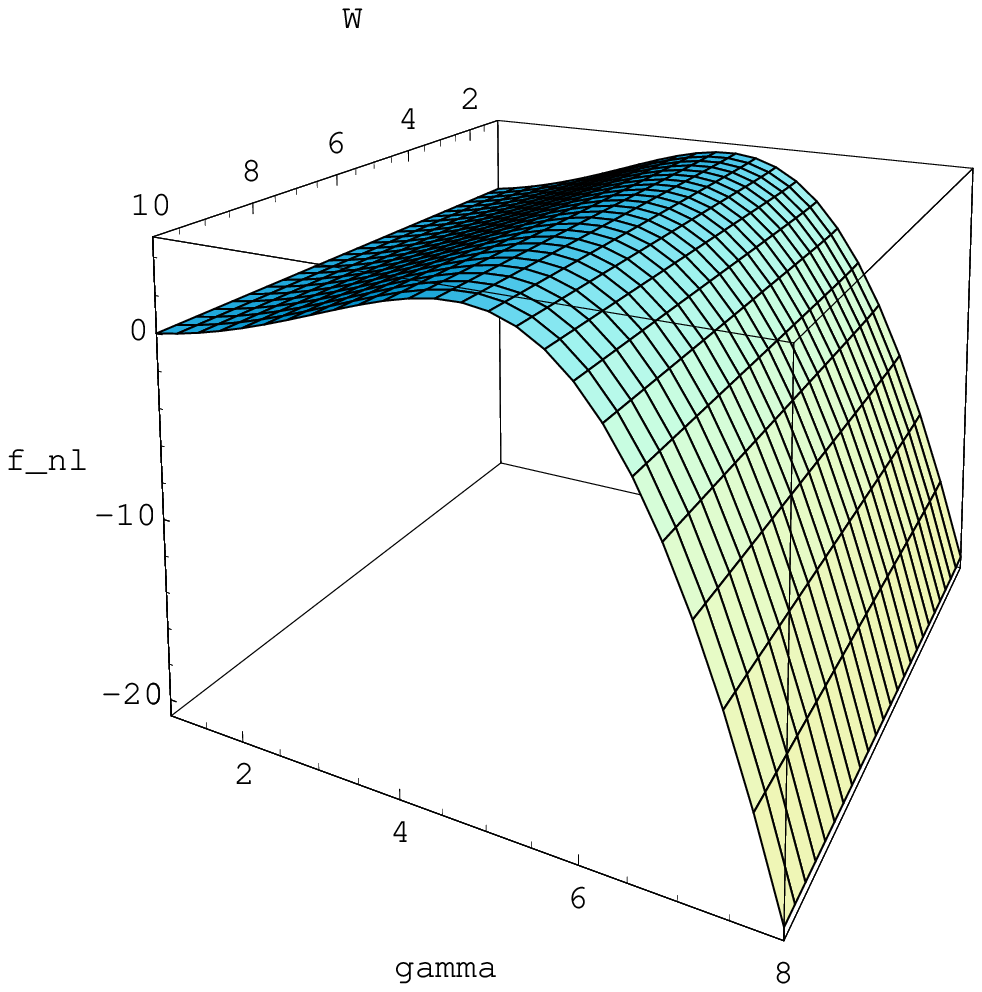}
\centering \texttt Figure 2: $N=50$
\end{minipage}
\begin{minipage}[t]{.50\textwidth}
\includegraphics[width=\textwidth]{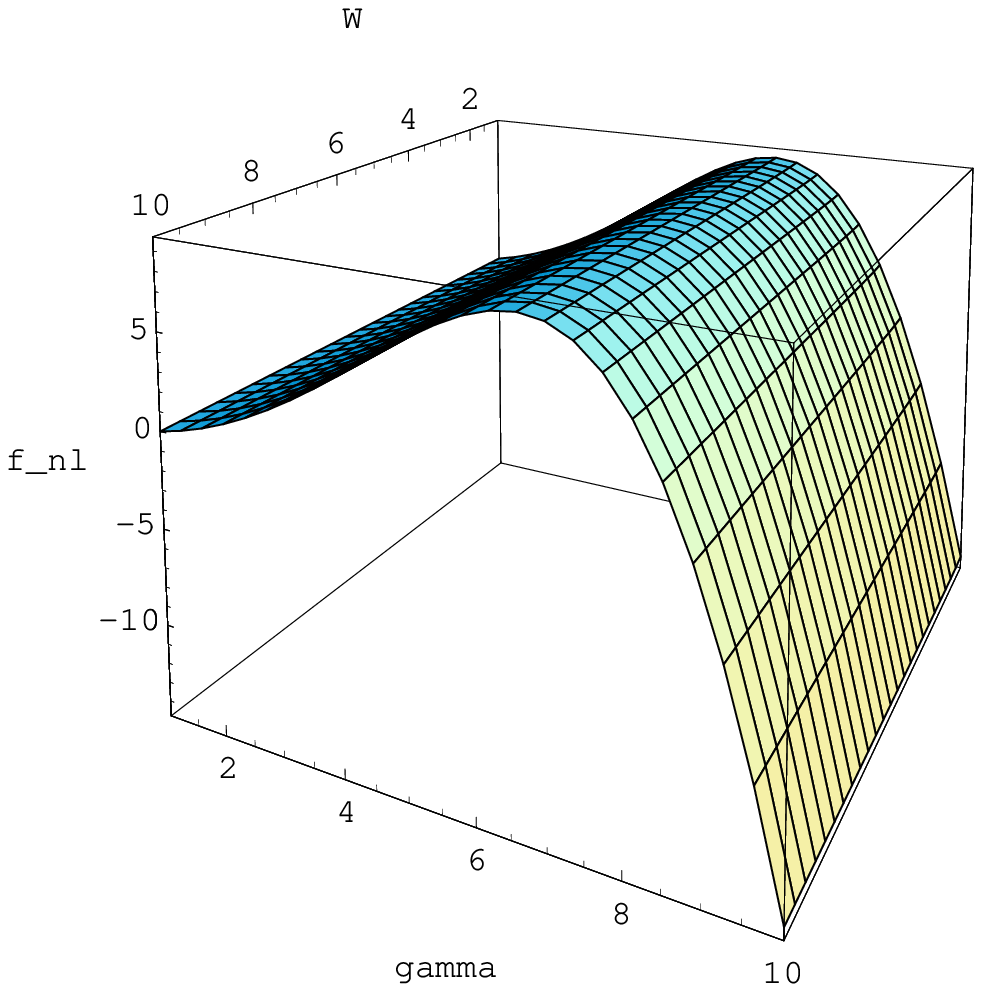}
\centering \texttt{Figure 3: $N=100$}
\end{minipage}
\begin{minipage}[t]{.50\textwidth}
\includegraphics[width=\textwidth]{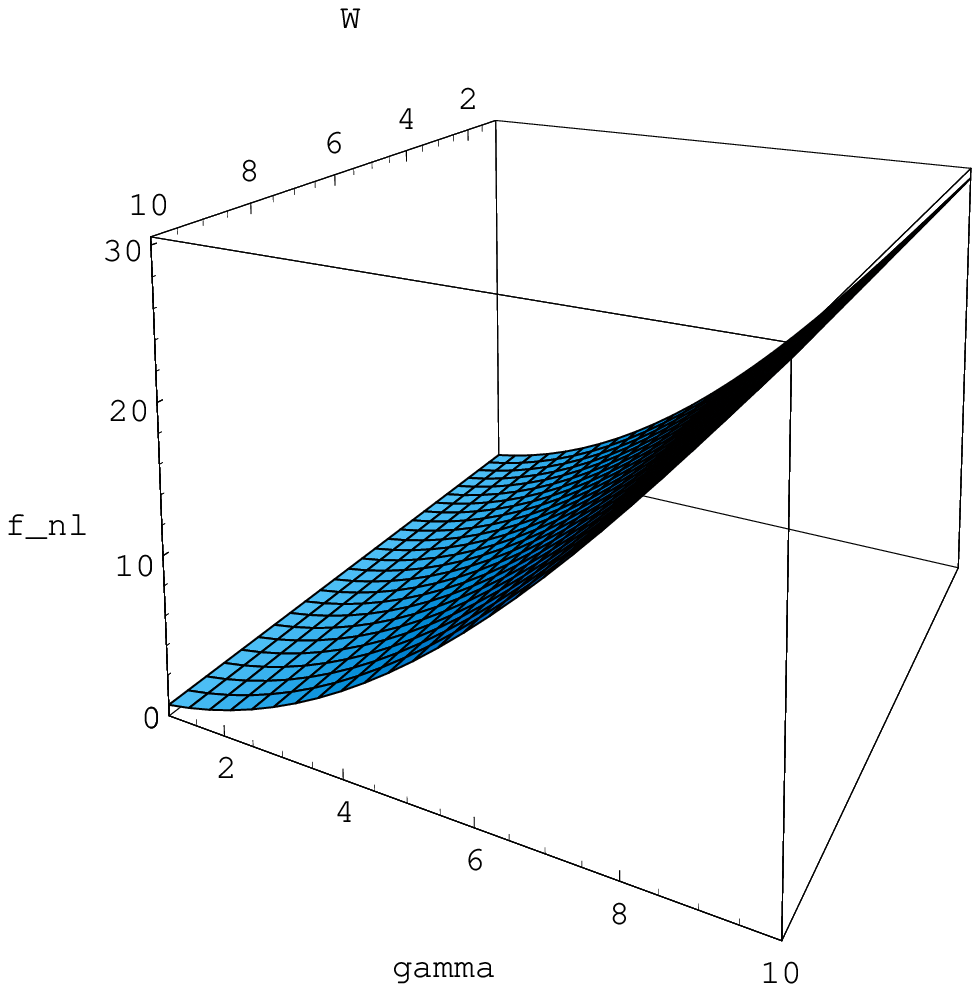}
\centering \texttt{Figure 4: $N=500$}
\end{minipage}
\end{figure}
\end{center}
\subsection{The limit of AdS warping.}
Another interesting case to consider is that of the $AdS_5 \times X^5$ solution, where the warp factor
goes like $h\sim \phi/(\sqrt{T_3} R)$. It is easy to see that $\alpha = -1, \beta =0$ in this instance and the resulting corrections to the $F_i$ functions
can be written below as (assuming that the potential is roughly constant over the regime of interest)
\begin{eqnarray}
F_1 &=& \frac{\gamma}{6 C_2}\left(\frac{(W^2-1)(1+3W^2)}{W^3}+\frac{2(W^2-1)(\gamma^2-1)}{W}+W(\gamma^2-1)(1+3\gamma^2) +\frac{24 V}{\gamma h^4}\right) \nonumber \\
&+& \frac{8\gamma}{6 C_2}\left(-\frac{(W^2-1)}{W}-W(\gamma^2-1)-\frac{4V}{\gamma h^4} \right) + \frac{8}{6C_2}\left(W\gamma -1 + \frac{V}{h^4} \right) \\
F_2 &=& \frac{1}{6\gamma C_2}\left(\frac{(W^2-1)(1+3W^2)}{W^3}-\frac{2(W^2-1)(\gamma^2-1)}{W}-W(\gamma^2-1)(3\gamma^2+1)+\frac{24\gamma V}{h^4} \right) \nonumber \\
&+& \frac{8}{6\gamma C_2}\left(-\frac{W^2-1}{W}+ W(\gamma^2-1)-\frac{4\gamma V}{h^4} \right) + \frac{8}{6C_2}\left(\frac{W}{\gamma}-1+\frac{V}{h^4} \right) +
\frac{W}{3\gamma C_2}(\gamma^2-1)^2. \nonumber
\end{eqnarray}
This expression is far more complicated than the solution in the constant background due to the explicit contribution from the warp factor. Note that 
the corrections in this case are now also dependent on the inflationary scale - due to the appearance of $V$ in the above expressions.
The fuzzy potential is now also a constant, given by $W_{AdS} = (1+4R^4/(\lambda^2 C_2))^{1/2}$, which explicitly depends on the ratio $M/N^2$

The resulting expression for the sound speed becomes
\begin{equation}
C_s^2 = \frac{1}{\gamma^2}\left(\frac{W^4(-4\gamma^2 + 3\gamma^4 - 6C_2)-2W^2(\gamma^2-1)-1}{W^4(8-24\gamma^2 + 15\gamma^4 - 6C_2)+W^2(8-6\gamma^2)-1} \right)
\end{equation}
which is again independent of both warp factor and potential. Note that when we take $\gamma$ to dominate, the sound speed approaches zero
much like in the large $N$ limit. Similarly taking $N \to \infty$ also reproduces the usual result proportional to $1/\gamma^2$. 
Unlike in the previous case, the sound speed now has zeros located at the following critical values of $\gamma$
\begin{equation}
\gamma^2_c = \frac{2}{3} + \frac{1}{3 W^2} \pm \frac{\sqrt{2}}{3 W^4} \sqrt{2W^4 - 4 W^6 + 2 W^8 + 9C_2 W^8}.
\end{equation}
The function becomes imaginary in between these zeros and should therefore be regarded as being an un-physical region of phase space. The width
of this region decreases as we increase $N$, and is therefore more pronounced for smaller values of $N$. For intermediate values of $N$, the
unphysical region is small, and the sound speed is small over a large range of $\gamma$.

If the fuzzy potential is taken to be large,
which is the most likely scenario due to the dependence on the flux/brane ratio, then it drops out of the expression altogether and we
are left with a two-parameter system - however the zeros of the function remain.

The effect of the zeros on the non-Gaussian amplitude are obvious, they give rise to singular spikes in the ${N, \gamma}$ phase space
which are located at larger values of $\gamma$ as we increase $N$. This is shown in Figure 5, where we assumed $W\sim10$ - although numerically
the value of $W$ has little effect on the overall behaviour until it becomes very large. Away from these spikes, the amplitude is always increasing monotonically
as one would expect since the corrections are washed out and shown in Figures 6-8. Physically the spectrum implies that the running of $f_{nl}$ with $\gamma$
is bounded, either from above or below once the $1/N$ corrections come into play. In the solution with large $N$, the singular region occurs
at very large values of $\gamma$ which already lie outside the observed experimental bounds.

Because the warping is not constant in this background, the additional field dependence induces a contribution to the running of the spectral index. 
In the ultra-relativistic approximation (which is unfortunately the only case that admits an analytic solution)
we find that
\begin{equation}
n_{nl} \sim 1+ \frac{4\gamma}{\phi_*} \sqrt{\frac{3 M_p^2}{N V(\phi_*)}}
\end{equation}
where $\phi_*$ denotes the field at horizon crossing. Since inflation also demands that the potential term dominates the kinetic term, this running
should still be small regardless of the precise form of the potential. Inflationary trajectories obey a similar slow roll expression as the one
in the constant warping limit, therefore implying that provided one tunes the fluxes and the potential inflation will be generic.

We have seen in this section how the subleading corrections distinctly alter the sign of the non-Gaussianities. In particular we note that
the $1/N$ correction leads to a maximal bound for the amplitude in constantly warped backgrounds 
whilst imposing restrictions upon the size of the parameter space in $AdS_5 \times X_5$ backgrounds.
Although both models are characterised by  $|f_{nl}| >>1$, and thus may satisfy the bounds (\ref{eq:ngbound}), the range of validity is 
restricted once we include $1/N$ corrections.
This is similar to the case examined in \cite{assistedinflationdbi} where the isocurvature perturbations lead to an unusual sign change relative
to the standard expression. 
This suggests that in general corrections to the large $N$ DBI inflation model 
will typically lead to new or refined signatures, which can be used as a more robust test of inflation in string theory. This is especially evident
for the finite $N$ case \cite{ward} where $N=2, 3$, since there the backreaction is fully under control but the sound speed runs like $C_s^2 \sim 1/(3 \gamma^2)$
which is even more suppressed relative to the single brane case, and therefore the non-Gaussian amplitude is \emph{enhanced}.

\begin{center}
\begin{figure}
\begin{minipage}[t]{.50\textwidth}
\includegraphics[width=\textwidth]{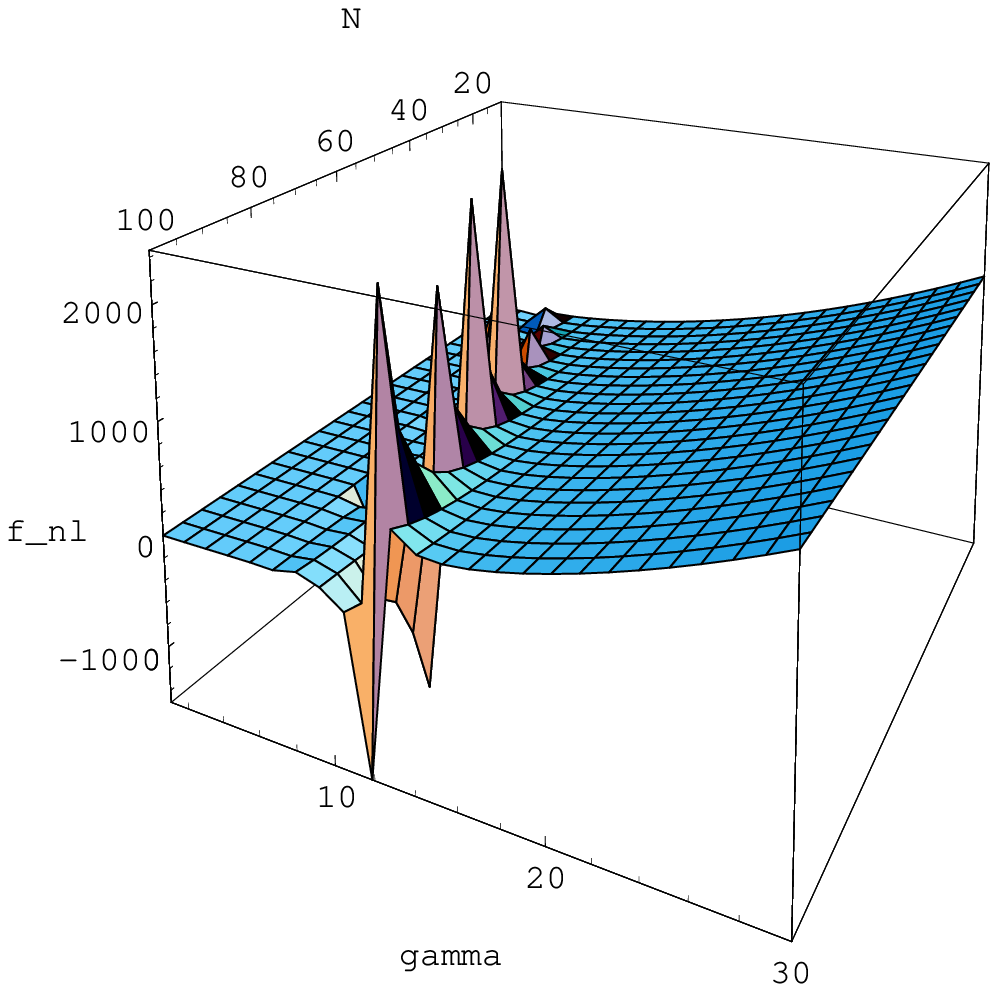}
\centering \texttt{Figure 5: Non-Gaussian amplitude in the $(N, \gamma)$ phase space, where $W=10$.}
\end{minipage}
\begin{minipage}[t]{.50\textwidth}
\includegraphics[width=\textwidth]{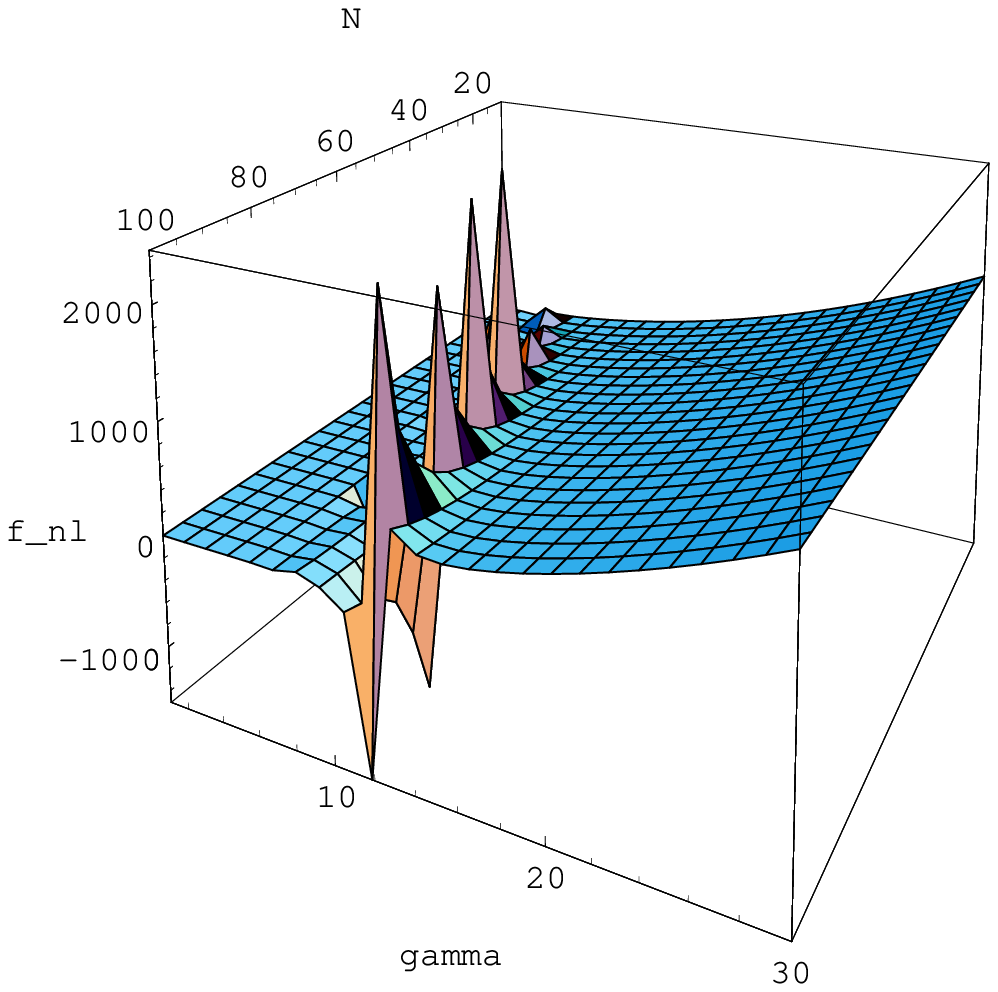}
\centering \texttt{Figure 6: Amplitude at $W=100$.}
\end{minipage}
\begin{minipage}[t]{.50\textwidth}
\includegraphics[width=\textwidth]{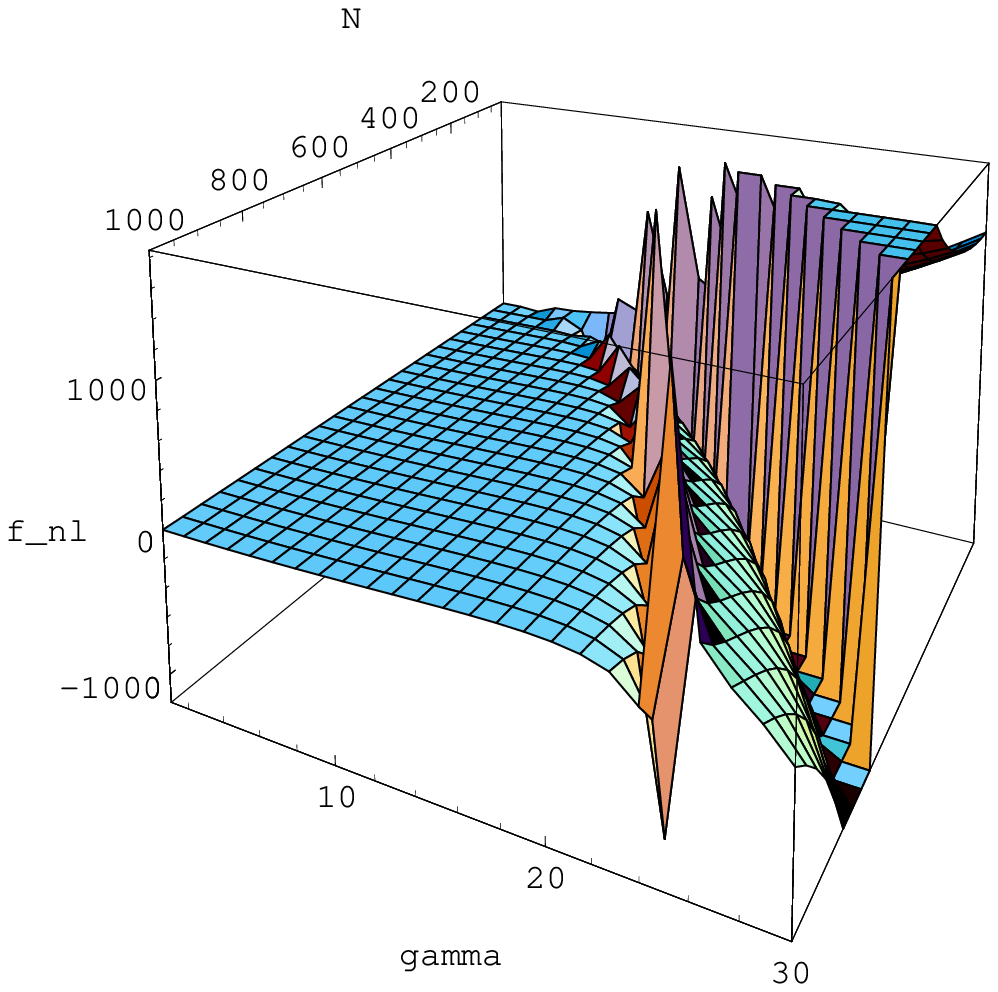}
\centering \texttt{Figure 7: Amplitude at $W=1000$.}
\end{minipage}
\begin{minipage}[t]{.50\textwidth}
\includegraphics[width=\textwidth]{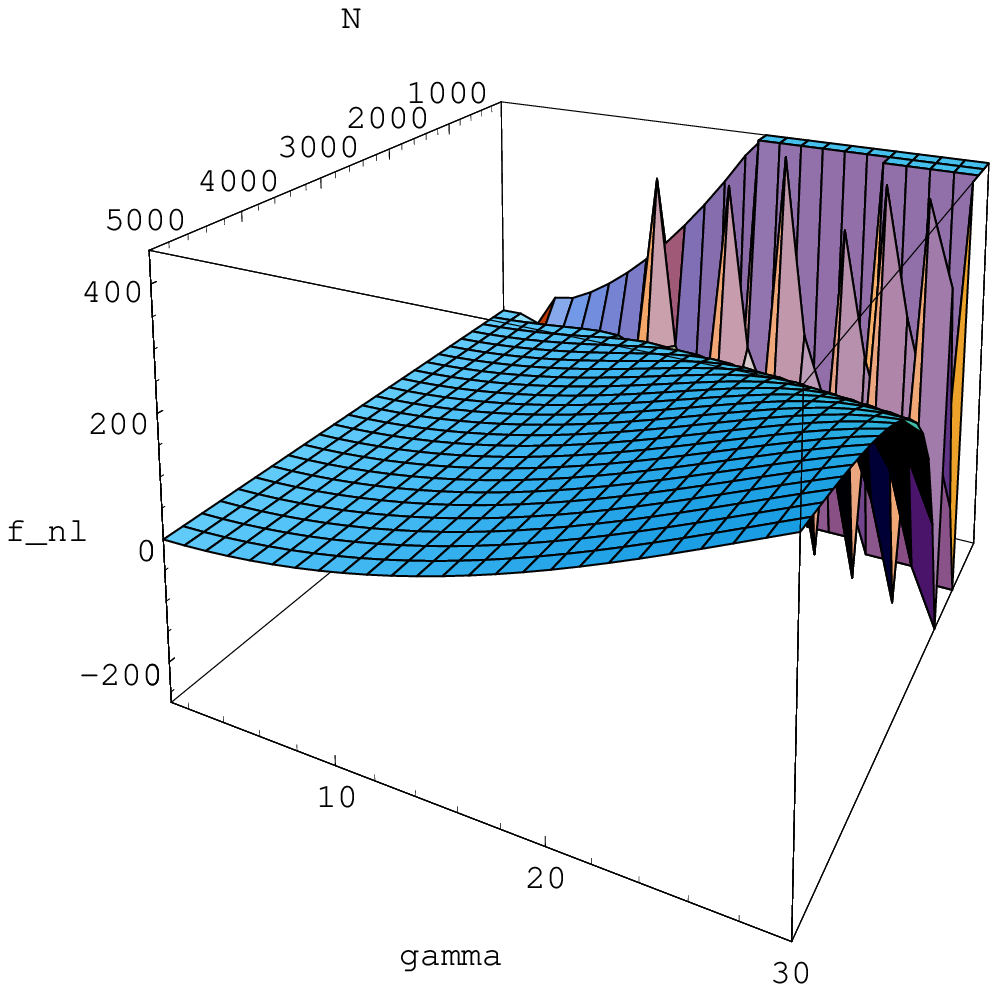}
\centering \texttt{Figure 8: Amplitude at $W=1000$, but with larger range of $N$.}
\end{minipage}
\end{figure}
\end{center}

\newpage
\section{Inflation via the Representation Cascade}

The analysis of DBI inflation using multiple $D3$-branes, or a wrapped $D5$-brane, allows us to evade many of the theoretical (and observational) constraints
present in inflationary model building. However besides the backreactive effects, there is also the issue of the scalar potential for the inflaton field.
Up to this point we have been rather cavalier about this, and just assumed that the potential is generated in the usual manner through interactions with
$\bar{D3}$-branes, $D7$-branes or symmetry breaking effects present in the compactification \cite{dbiinflation}. 
Whilst this is a reasonable assumption for the $D3$-brane
scenario, one must ask about the validity of this for the case of wrapped branes \cite{wrappedbranes}. 
In the absence of a pure string calculation, we are simply forced
to insert the scalar potential by hand. However the non-Abelian structure of the Myers action already contains a potential term, and so one could
enquire whether this could be used to drive a period of inflation \cite{myers}.

Because the world-volume theory for coincident branes is non-Abelian we find that the induced scalars are no longer singlets,
but are instead promoted to matrices. The simplest solution for all these problems is to use the symmetry of the transverse space
and select the scalars to lie in representations of $SO(6)$. A particularly nice and simple choice, which is the one that gives us a theory
dual to a spherically wrapped $D5$-brane, is to assume the scalars are valued in $SO(3) \sim SU(2)$. Typically we represent this through
the ansatz
\begin{equation}
\phi^{i} = \hat{R}\alpha^i
\end{equation}
where $\alpha^i$ are the $N$-dimensional irrep generators of $SU(2)$ and $\hat R$ is some parameter with dimensions of mass.
Because our theory is embedded in a non-commuting target space, we see that our geometry is also non-commuting and because
of the identification with $SO(3)$ we argue that the scalars lie on a fuzzy $S^2$, the radius of which is defined by
\begin{equation}
r^2 = \frac{\lambda^2}{N} \rm{Tr}(\phi^i \phi^i) = \lambda^2 \hat{R}^2 C_2
\end{equation}
where as usual we denote the Casimir of $SU(2)$ by $C_2$.

Previous work on brane inflation has implicitly assumed that the open string mode $r$ should be identified as the inflaton, 
leading to a sustained period of inflation. The change of the warp factor as a function of $r$ allows for DBI inflation to occur in these warped models.
However this justification already assumes that the scalar field is in the irreducible representation of the gauge group. A priori
there is no reason why this representation will be selected, and so one could argue for solutions where the initial configuration
was a reducible representation. Of course we know that the irrep will lead to the smallest energy configuration, so we would expect
the reducible solution to cascade down to the irrep. The 
representation flow itself will appear as a scalar field on the world-volume and could therefore be an inflaton candidate.
Indeed this would seem to be the most generic behaviour given the context of the string landscape.

So our primary assumption in this section is that the branes are static and fixed at some point in the IR of a warped throat. But the scalar fields
are now initially in a reducible representation. How do we model such a cascade? A simple example. which should be
representative of a more general class of solutions, is to take as an ansatz \cite{jatkar, n=1*}
\begin{equation}
\phi^i = \hat{R}\left((1-g(t))\alpha^i + g(t) J^i \right)
\end{equation}
where now $J^i$ is a generator in a reducible representation of the gauge group, and we fix the boundary conditions as $g(0)=1$
and $g(t_e)=0$, so that as time evolves the scalar flows from the reducible to the irreducible representation.
Dynamical transitions such as these occur in a class of $\mathcal{N}=1^{*}$ SYM theories, where the choice of representation
has important physical properties. Since our branes
are not dynamical we see that $\hat{R}$ is independent of time.
A nice, and convenient parameterisation, is to choose $J^i$ such that it corresponds to the spin $j^{\prime} \oplus j^{\prime}$ representation i.e the reducible
representation is comprised of two blocks of $N^{\prime} = N/2$. 
Physically this means that our moduli space consists of two coincident fuzzy spheres which coalesce to form a single
sphere \cite{jatkar}. To further simplify things we denote $D_2$ as the Casimir of the reducible rep $D_2 = N^2/4-1$, and we will also assume that 
$\rm{Tr}(\alpha^i J^j) = 0$.

Plugging all this into the coincident $D3$-brane action, for large $N$, we see that it can be written as follows
\begin{eqnarray}
S &=&- N T_3 \int d^4 \xi h_c^4 \sqrt{1-\frac{r^2(C_2+D_2) \dot{g}^2}{h_c^4(C_2(1-g)^2+g^2D_2)}}\sqrt{1+\frac{1}{h^4 \lambda^2}\frac{4r_c^4}{(C_2(1-g)^2+g^2D_2)}} \nonumber \\
&=& - N T_3 \int d^4 \xi h_c^4 (W(r,g) \tilde \gamma^{-1}-1)
\end{eqnarray}
where $C_2$ is once again the effective Casimir of the irrep, whilst $D_2$ is the effective Casimir of the reducible representation and 
$W(r,g)$ is the fuzzy potential. We have also included the contribution from the Chern-Simons term in the last line above, and we have denoted the 
fixed value or $r$ by $r_c$. Finally $\tilde \gamma$ is the obvious analogue of the kinetic contribution to the action - not to be confused with $\gamma$ in the previous
sections.
In general the full inflationary dynamics will depend on both $g,r$ as the branes move through the throat, and also
undergo the cascade in representation space. However the analysis is complicated. Note that we are not making any assumptions about the specific background here, all that 
we are imposing is that the radial term is approximately constant during the cascading phase.

One can see that as $t \to t_e$ the theory reproduces the action in the first chapter as it should.
We now couple this action to four-dimensional Einstein gravity. Because we are assuming that the branes are not moving in the 
radial direction, then both the Chern-Simons term and any scalar potential we can add to the action will simply be constants and will drop out of the dynamical analysis, thus essentially we have a theory reminiscent of the open string tachyon dynamics - where the only terms of interest
arise through the $NS$-sector alone.
Analysis of the static potential shows that (aside from the boundary conditions) there is a local maximum at $g_c= C_2/(C_2+D_2)$, which
approaches $4/5$ as $N$ increases. This is in fact a tachyonic point of the theory, but is smoothed out somewhat once we turn on velocity terms. 
This provides a small barrier for the field as it rolls towards $g=0$. If the field has no
initial kinetic energy - then the inflaton will sit near $g=1$ and the energy density (hence Hubble parameter) will 
essentially be constant and can drive a sustained period of inflation. The field can tunnel through this barrier, and will
eventually flow towards its boundary point - which is indeed the lowest energy configuration as we argued for.
This is essentially a phase of 'Old Inflation', although in a new context.

We see that the following general cosmological equations must be satisfied
\begin{eqnarray}
H^2 &=& \frac{N T_3 h^4 (W \tilde \gamma - 1)}{3 M_p^2}\nonumber \\
\frac{\ddot a}{a} &=& \frac{N T_3 h^4 }{3 M_p^2}\left(\frac{3W}{2\tilde \gamma}-\frac{W \tilde \gamma}{2} \pm 1 \right)
\end{eqnarray}
where the $+$ sign corresponds to a $\bar{D}3$-brane, whilst the minus sign is for the usual $D3$-brane.
Typically in DBI inflation we assume the existence of $D3$-branes. For IR inflation this appears natural since one can assume that
the initial conditions conspire to create $D3$-branes after flux annihilation. For UV models, we are assuming that the scalar potential
is generated by $\bar{D}3$-branes which sit in the IR of a throat to screen the relative $D3$-charge of the RR fluxes. Therefore
in order to realise a solution where we have coincident $\bar{D}3$-branes, we could assume that they are positioned at some point
in the throat as a screen for the fluxes. The uplifting of Minkowski vacua in the simplest KKLT scenario \cite{kklt} employs such a configuration
(albeit with a single brane), however the ISD nature of the fluxes prevents the branes from being dynamical objects\footnote{Although these terms explicitly break the $\mathcal{N}=1$ supersymmetry of the solution}. So this configuration is not as unnatural as it may first appear.

From these expressions it follows immediately that inflation is only possible when the following inflationary constraint is satisfied
\begin{equation}\label{eq:irrepconstraint}
W(3- \tilde \gamma^2) \ge \mp 2 \tilde \gamma \hspace{0.5cm} \to \hspace{0.5cm} \tilde \gamma \le \frac{\sqrt{1+3 W^2} \pm 1}{W}
\end{equation}
where in the last step we wrote this explicitly as a constraint upon $\tilde \gamma$. Note that in the limit of extremely large $W$ satisfying $W>>1$, this condition 
reduces to the non-relativistic approximation that $\tilde \gamma^2 \le 3$ for both types of branes. 
Note that these two limits are compatible with each other and therefore there is a small 
inflationary window available.
Note the importance of the Chern-Simons term above.
 
For $D3$-branes we see that when $W\sim 1$ the constraint collapses to $\tilde \gamma = 1$ which implies
that the solution is non-dynamical. Clearly for this to be a physical solution we are forced to fix the field at $g=1$ which is the location of the 
metastable de-Sitter minimum. Thus inflation will occur for as long as the system is in the reducible representation. Eventually the field must tunnel
out from this false vacuum via the Hawking-Moss instanton\footnote{This is because the barrier separating the two minima is relatively small in height, 
and thus the no-wall approach provides a better description than the thin wall approximation \cite{hawkingmoss, kklt}.} and inflation will end rapidly.
Let us estimate the probabilities and associated time scales for this to occur. First it is necessary to re-write the action in canonical form which can
be achieved through the following field re-definitions
\begin{equation}
V(\phi) = N T_3 h_c^4 (W-1), \hspace{0.5cm} \phi = \sqrt{2 N T_3 h_c^4 r^2  (C_2+D_2)} \int \frac{\sqrt{W}dg}{\sqrt{C_2(1-g)^2+g^2D_2}}.
\end{equation}
As the potential can be seen to vanish when $W=1$, we must consider the slow-roll expansion of the DBI action in order to derive the above
conditions. If we allows the inflaton to have a small, but non-zero velocity, then this automatically forces $W >1$ and a potential exists.
Now the Hawking-Moss instanton solution treats the inflaton as undergoing Brownian motion from the false vacuum to the global maximum. The tunneling suppression
probability  is given by
\begin{equation}
P=\exp\left(-\frac{24 \pi^2 M_p^4}{V(\phi_0)}+\frac{24 \pi^2 M_p^4}{V(\phi_1)} \right)
\end{equation}
where the false vacuum is defined at $V(\phi_0)$. The result of Hawking and Moss is that because of the Brownian motion there is not homogeneous tunneling, 
rather the homogeneity is spread over the scale $H^{-1}$.
With our solution the tunneling probability is therefore well approximated by
\begin{equation}
P \sim \exp\left(-\frac{12 \pi^2 M_p^4 \lambda^2 D_2^2}{N T_3 r^4(C_2+D_2)}\right) 
\end{equation}
where the exponent runs like $N$ with our specific choice of representations. Therefore the term in the exponent is large and negative, indicating that
the tunneling suppression probability is relatively small. This means that the field will most likely tunnel through the small barrier than climb over it.
The time for decay should therefore also be small and is given by
\begin{equation}
t_{decay} \sim t_r \exp \left(-\frac{12\pi^2 M_p^4 \lambda^2 C_2 D_2}{ N T_3 r^4(C_2+D_2)} \right)
\end{equation}
where $t_r$ is the recurrence time defined through the relation $t_r \sim \exp(24 \pi^2 M_p^4/V(\phi_0))$. Clearly $t_{decay} << t_r$ indicating that
the solution will quickly tunnel from the false vacuum. Again the solution is exponential decreasing as a function of $N$ with our representation
choice.

The fact that the inflaton rolls to its global minimum does offer the possibility of avoiding the graceful exit
problem that plagues models of old inflation. For all other solutions, the field must be in a phase of slow roll satisfying $1 \le \tilde \gamma^2 \le 3$
which is a very restrictive condition. Because of the field dependence of both potential and kinetic terms, one would expect this inflationary phase to
end rapidly - yielding \emph{at the very most} a single e-folding of inflation. If one views this optimistically then it may be possible to obtain the
requisite amount of e-foldings by assuming that there are at least 60 reducible representations. In this case the field will roll down a potential, which is rather
step like in shape - each transition contributing a single e-folding. From the non-commutative geometry viewpoint this would be interpreted as the steady
coalescence of $N/60$ fuzzy spheres into a single sphere. 
Indeed one could model such a flow by reverting to an ansatz of the form
\begin{equation}
\phi^k = \hat{R} \left( \sum^d_{\alpha=0} N_{\alpha} g_{\alpha}(t) J^{k \alpha} + (N_{\alpha} - g_{\alpha}(t))J^{k (\alpha+1)}\right)
\end{equation}
where $N_{\alpha}$ is an appropriate boundary parameter for the flow through representation space, and we are summing over representations from $0\ldots d$,
with the zeroth representation being the fundamental one.

One may argue that this is not the most general dynamical process, since several spheres may coalesce at the same time and thus 
prevent the system from generating enough inflation. One may also argue that this would typically require $N$ to be much larger than originally presumed in
order for there to be such a large number of reducible representations. This poses a problem since the back-reaction will inevitable be uncontrollable.
However one possible resolution to these problems lies in the fact that the model is extremely simple, relying on the fact that the brane stack
is fixed in spacetime. This will not be the most general solution, and in fact we expect both the representation cascade and the open string modes associated
with the radial embedding will combine to drive inflation. This is analogous to a purely spinflation based model (see the nice discussion of this effect in \cite{spinflation}), 
where the branes are fixed at some radial distance in the throat geometry but have non-trivial angular momentum. 
The amount of inflation obtained is roughly the same in both cases.

In the case of $\bar{D}3$-branes with $W \sim 1$ we find that $\tilde \gamma^2 \le 9$ which is a much weaker constraint
on the flow velocity compared to the $D3$-brane case - although it falls into the region of 'intermediate' velocities
i.e somewhere between slow roll and relativistic rolling. In fact it can be seen that the maximal allowed value of $\tilde \gamma$ is a decreasing function
of the fuzzy potential. Let us consider this solution. Assuming that the velocity saturates the bound on $\tilde \gamma$, and using the continuity
equation we find that (up to a factor of $1/\tilde \gamma^2 \sim 1/9$)
\begin{equation}
\dot{g} \sim -\frac{16 M_p^2 H' F(g)}{9 N T_3 \tilde \gamma r^2 (C_2+D_2)}
\end{equation}
where we have defined $F(g) = C_2(1-g)^2 + D_2 g^2$ as the flow parameter. Solving for $\tilde \gamma$ as a function of $g$ 
we can then estimate the primary slow roll parameter 
to be
\begin{equation}
\epsilon_1 \sim \frac{h^2 \sqrt{F(g)}}{C_2+D_2} \left(\frac{H'}{H} \right)^2.
\end{equation}
Since this is suppressed by the warpfactor, and the largest possible value of the remaining prefactor is $D_2/(C_2+D_2)$ which is less than unity, the
Hubble terms are the most important. At leading order we then find the following bound on the Hubble scale during inflation
\begin{equation}
H^2 > \frac{9 \times 10^{-2} M_p^4 h_c^2 (5N^2-4)\sqrt{F(g)}}{r_c^2}
\end{equation}
where the lowest Hubble scale occurs around $g\sim g_c$ as one would expect. This will also set the scale of the 
tensor perturbations since they are proportional to $H^2$ at horizon crossing. As one would anticipate, the Hubble scale increases with the number
of branes, but is still modulated by the warp-factor. For solutions such as Klebanov-Strassler, the warp factor is exponentially suppressed at the tip of the 
throat and therefore the Hubble scale will be lower in throats with large flux quanta.

The equation of motion for the inflaton is given by
\begin{equation}
\frac{\dot{g}^2 \alpha \beta }{2} - 3H \alpha \dot{g} - V' \sim 0
\end{equation}
where we have dropped terms proportional to $\ddot{g}$ as is usual for slow roll models. To simplify the expression
we have used the following definitions
\begin{equation}
\alpha = \frac{2N T_3 h_c^4 r_c^2 W(C_2+D_2)}{C_2(1-g)^2+D_2g^2}, \hspace{0.5cm} \beta = 1-\frac{2 \alpha'}{\alpha}
\end{equation}
and primes denote derivatives with respect to the inflaton.
For solutions where $\dot{g}^2 << (W+1)(C_2 (1-g)^2+g^2D_2)/(W r^2 (C_2+D_2))$, we can neglect the kinetic contribution to the Hubble parameter and therefore
we can explicitly solve for the inflaton to find
\begin{equation}\label{eq:eom}
\dot{g} \sim \frac{\sqrt{3 V}}{\beta M_p}\left(1 \pm \sqrt{1-\frac{\beta M_p^2 r^4 F'}{3 h^4 \pi^2 l_s^4 \alpha W(W+1)F^2}} \right)
\end{equation}
which is a complicated function of the flow parameter. Numerically we can scan the space of solutions, and we see that inflation 
is possible but only a handful of e-foldings are generated. This suggests that our simple model must be modified in order for it
to be a viable candidate. There are at least two ways in which this could occur. Firstly as already mentioned, we can allow for
the field to be in a different initial representation so that the cascade has more steps. If there are $n$ different transitions
each yielding $N_e$ efolds of inflation then we may anticipate that the model could generate $n N_e$ e-foldings during 
the cascade. This could easily be tuned to satisfy the WMAP data \cite{data}. The alternative is to consider this as a multi-field model
where the inflaton is some combination of $g, r$ where $r$ represents the radial motion in the throat. Indeed our assumption that the
branes are fixed is in principle difficult to achieve due to interactions with the fluxes. Therefore we could generally
that the combined amount of inflation driven by dynamical branes and also by the cascade, will easily satisfy the bounds. Moreover
the constraints on the brane positions will be slightly weakened due to the presence of the extra fields.

\section{Discussion}
Cosmology has entered into a new era of precision data \cite{data}, and it is therefore imperative that top-down models make some falsifiable predictions in order to
distinguish them from simple field theory phenomenology. The simplest models in string theory belong to the class of DBI inflation, and their defining
characteristic is that the sound speed of fluctuations is greatly suppressed leading to potentially observable non-Gaussian signatures in the CMB \cite{dbiinflation}.
Whilst this has been an important result, recent work has determined that we need to develop more intricate models in order to completely
satisfy the current observational data. Extending these scenarios to include wrapped branes, or multiple branes allows us to evade these constraints at
the cost of losing control over the low energy theory. The previous work \cite{ward} demonstrated new physical effects when one uses a finite number of 
coincident branes, this is the best of both worlds in some sense - since we can still control the backreaction of the branes upon the warped 
geometry, but also capture more interesting world-volume effects. In this paper we have investigated the effect of the $1/N$ corrections to the 
large $N$ solution, since this is essentially a combinatorics issue \cite{finiten} and also overlaps with much of the recent work on wrapped configurations \cite{wrappedbranes}.
Specifically we have seen how these corrections, suppressed in the large $N$ limit, affect the speed of sound and the non-Gaussianity (at least for 
the equilateral triangle modes) in backgrounds with constant warping, and backgrounds of the form $AdS_\times X_5$. In both cases we have seen that the
spectrum of non-Gaussianities has new features present in the non-relativistic (constant warping) and intermediate ($AdS_5$) limits.
This indicates that the $1/N$ correction plays an interesting role in the inflationary dynamics.
We have also started to develop an alternative inflationary scenario using a cascade through representation space to drive inflation. 
This appears to be sensitive to the charge of the $D3$-brane and also imposes tight restrictions on the inflaton velocity. 
For the usual $D3$-brane solution, we find that the field prefers to tunnel from
the false vacuum to the true vacuum and that the decay rate for such a vacuum is relatively short and goes like $e^{-N}$ at large $N$.
This is very much reminiscent of old inflation, although because the field will still roll towards the minimum it may evade the graceful exit problem.
For $\bar{D}3$-branes on the other hand, slow roll inflation appears to be preferred - although we estimate that in order for the model to satisfy the 
COBE bounds we require multiple transitions which may not be feasible.

There remains much work to do on building viable models of DBI inflation. The results shown here and in \cite{assistedinflationdbi} have shown that
the non-Gaussianities can be significantly different from the leading order term once you start to include sub-leading effects. This suggests that
other corrections could also become important, even in the slow roll regime \cite{slowroll} of DBI inflation. These corrections could be 
particularly interesting for the finite $N$ solutions in \cite{ward}, since the backreaction of the branes on the warped geometry is still under
control - however the sound speed has dramatically different behaviour to the large $N$ and single brane models. 
We must also develop better mechanisms for reheating \cite{reheating} in these models. Since the non-linear form of the action captures all the terms in the $\alpha'$
expansion, one would hope that there could be some stringy signature present in the standard model which lies just beyond the current collider physics scale.
This is important not only for aesthetics, but also because the string signature can again be tested.
We hope to return to these issues in the future.

\begin{center}
\textbf{Acknowledgements}
\end{center}
We wish to thank Steve Thomas, Larus Thorlacius, Adam Ritz, Shinji Tsujikawa, Shinji Mukohyama and Maxim Pospelov for useful remarks and comments.


\begin{thebibliography}{99}
\bibitem{lectures}
C. P. Burgess, 0708.2865 [hep-th]; L. McAllister and E. Silverstein, arXiv:0710.2951 [hep-th], J. M. Cline, hep-th/0612129; A. Linde, Contemp. Concepts Phys {\bf 5}, 1-362 (2005), hep-th/0503203, A. Linde, J. Phys. Conf. Ser {\bf 24}, 151-160 (2005), hep-th/0503195.
\bibitem{data}
D. N. Spergel et al, ApJS {\bf 170}, 377 (2007), astro-ph/0604449;
S. Perlmutter et al, Astrophys. J. {\bf 517}, 565-586 (1999), astro-ph/9812133;
A. G. Riess et al, Astron. J. {\bf 116}, 1009-1038, astro-ph/9805201; 
A. G. Riess et al, astro-ph/9810291;
W. J. Percival et al, Mon. Not. Roy. Astron. Soc. {\bf 327}, 1297 (2001) astro-ph/0105252;
S. Cole et al, Mon. Not. Astron. Soc. {\bf 362}, 505-535 (2005), astro-ph/0501174;
M. Tegmark et al, Phys. Rev. {\bf D 73}, 023502 (2006), astro-ph/0608632.
\bibitem{darkenergy}
E. J. Copeland, M. Sami and S. Tsujikawa, Int. J. Mod. Phys. {\bf D 15}, 1753-1936 (2006), hep-th/0603057.
\bibitem{geometricflux}
F. Denef, M. R. Douglas and S. Kachru, hep-th/0701050;
M. R. Douglas and S. Kachru, hep-th/0610102;
M. Grana, Phys. Rept. {\bf 423}, 91-158 (2006), hep-th/0509003;
S. B. Giddings, S. Kachru and J. Polchinski, Phys. Rev. {\bf D 66}, 106006 (2002), hep-th/0105097.
\bibitem{nongeomflux}
B. Wecht, arXiv:0708.3984 [hep-th];
S. Kachru, M. B. Shulz, P. K. Tripathy and S. P. Trivedi, JHEP {\bf 0303}, 061 (2003), hep-th/021182.
\bibitem{cosmicstrings}
E. J. Copeland, R. C. Myers and J. Polchinski, JHEP {\bf 0406}, 013 (2004), hep-th/0312067;
E. J. Copeland, P. M. Saffin, JHEP {\bf 0511}, 023 (2005);
M. Sakellaridou, hep-th/0602276;
H. Firouzjahi, L. Leblond and S.H. Henry Tye, JHEP {\bf 0605}, 047 (2006), hep-th/0603161;
S. Thomas and J. Ward, JHEP {\bf 0612}, 057 (2006), hep-th/0605099;
H. Firouzjahi, JHEP {\bf 0612}, 031 (2006), hep-th/0610130;
L. Leblond and M. Wyman, Phys. Rev. {\bf D 75}, 123522 (2007), astro-ph/0701427;
\bibitem{perturbations}
J. Garriga and V. F. Mukhanov, Phys. Lett. {\bf B 458}, 219-225 (1999), hep-th/9904176, J. Hwang and H. Noh, Phys. Rev. {\bf D 66}, 084009 (2002), hep-th/0206100; J. Hwang and H. Noh, Phys. Rev. {\bf D 71}, 063536 (2005), gr-qc/0412126.
\bibitem{realisticmodels}
D. Baumann, A. Dymarksy, I. R. Klebanov, L. McAllister and P. J. Steinhardt, arXiv:0704.3837 [hep-th];
S. Panda, M. Sami and S. Tsujikawa, arXiv:0707.2848 [hep-th];
J.J. Blanco-Pillado, C. P. Burgess, J. M. Cline, C. Escoda, M. Gomez-Reino, R. Kallosh, A. Linde and F. Quevedo, JEHP {\bf 0609}, 002 (2006); 
A. Krause and E. Pajer, arXiv:0705.4682 [hep-th].
\bibitem{braneinflation}
S. H. Henry Tye, hep-th/0611148;
S. E. Shandera and S. H. Henry Tye, JCAP {\bf 0606}, 011 (2006), hep-th/0602136;
S. Shandera, B. Shlaer, H. Stoica and S. H. Henry Tye, JCAP {\bf JCAP 0402}, 013 (2004), hep-th/0311207;
G. Shiu and S. H. Henry Tye, Phys. Lett. {\bf B 516}, 421-430 (2001);
E. Halyo, hep-th/0402155; G. Dvali, Q. Shafi and S. Solganik, hep-th/0105203.
\bibitem{dbarinflation}
S. H. S. Alexander, Phys. Rev. {\bf D 65}, 023507 (2002). hep-th/0105032;
C. P. Burgess, M. Majumdar, D. Nolte, F. Quevedo, G. Rajesh and R-J. Zhang, JHEP {\bf 0107}, 047 (2001);
C. P. Burgess, P. Martineau, F. Quevedo, G. Rajesh and R-J. Zhang, JHEP {\bf 0203}, 052 (2002), hep-th/0111025;
C. Choudhury, D. Ghoshal, D. P. Jatkar and S. Panda, JCAP {\bf 0207}, 009 (2003), hep-th/0305104.
\bibitem{dbiinflation}
E. Silverstein and D. Tong, Phys. Rev. {\bf D 70}, 103505 (2004), hep-th/0310221;
M. Alishahiha, E. Silverstein and D. Tong, Phys. Rev. {\bf D 70}, 123505 (2004), hep-th/0404084.
\bibitem{assistedinflation}
A. R. Liddle, A. Mazumdar and F. E. Schunk, Phys. Rev. {\bf D 58}, 061301 (1998), astro-ph/9804177;
P. Kanti and K. A. Olive, Phys. Rev. {\bf D 60}, 043502 (1999), hep-ph/9903254;
K. A. Malik and D. Wands, Phys. Rev. {\bf D 59}, 123501 (1999), astro-ph/9804177;
E. J. Copeland, A. Mazumdar and N. J. Nunes, Phys. Rev.{\bf D 60}, 083506 (1999), astro-ph/9904309;
J. Hartong, A. Ploegh, T. Van Reit and W. B. Wastra, Class. Quant. Grav {\bf 23}, 4593-4614, gr-qc/0602077;
K. L. Panigrahi and H. Singh, arXiv:0708.1679 [hep-th];
H. Singh, Mod. Phys. Lett. {\bf A 22}, 2737 (2007), hep-th/0608032;
H. Singh, Nucl. Phys. {\bf B 734}, 169 (2006), hep-th/0508101;
A. Mazumdar, S. Panda and A. Perez-Lorenzana, Nucl. Phys. {\bf B 614}, 101-116 (2001), hep-ph/0107058;
M. Majumdar and A. Davis, Phys. Rev. {\bf D 69}, 103504 (2004), hep-th/0304226;
Y-S. Piao, R-G. Cai, X. Zhang and Y-Z. Zhang, Phys. Rev. {\bf D 66}, 121301 (2002), hep-ph/0207143.
\bibitem{assistedinflationmtheory}
K. Becker, M. Becker and A. Krause, Nucl. Phys. {\bf B 715}, 349-373 (2005), hep-th/0501130;
J. Ward, Phys. Rev. {\bf D 73}, 026004 (2006), hep-th/0511079;
A. Ashoorion and A. Krause, hep-th/0607001;
A. Krause, arXiv:0708.4414 [hep-th];
E. I. Buchbinder, Nucl. Phys. {\bf B 711}, 314-344 (2005), hep-th/0411062.
\bibitem{nflation}
S. Dimopolous, S. Kachru, J. McGreevy and J. Wacker, hep-th/0507205;
R. Easther and L. McAllister, JCAP {\bf 0605}, 018 (2006), hep-th/0512102;
Y-S. Piao, Phys. Rev. {\bf D 74}, 047302 (2006);
M. E. Olsson, JCAP {\bf 04}, 019 (2007), hep-th/0702109.;
S. A. Kim and A. R. Liddle, 0707.1982 [astro-ph].
\bibitem{assistedinflationdbi}
M-x. Huang, G. Shiu and B. Underwood, arXiv:0709.3200 [hep-th].
\bibitem{multibrane}
J. M. Cline and H. Stoica, Phys. Rev. {\bf D 72}, 126004 (2005), hep-th/0508029.
\bibitem{ward}
S. Thomas and J. Ward, Phys. Rev. {\bf D 76}, 023509 (2007), hep-th/0702229.
\bibitem{constraints}
D. H. Lyth, Phys. Lett. {\bf 78}, 1861-1863 (1997), hep-ph/9606387;
D. Baumann and L. McAllister, Phys. Rev. {\bf D 75}, 123508 (2007);
J. E. Lidsey and I. Huston, JCAP {\bf 0707}, 002 (2007), arXiv:0705.0240 [hep-th].
\bibitem{wrappedbranes}
T. Kobayashi, S. Mukohyama and S. Kinoshita, arXiv:0708.4285 [hep-th];
S. Mukohyama, arXiv:0706.3214 [hep-th];
M. Becker, L. Leblond and S. Shandera, arXiv:0709.1170 [hep-th].
\bibitem{irinflation}
R. Bean, X. Chen, H. V. Peiris and J. Xu, arXiv:0710.1812 [hep-th];
R. Bean, S. E. Shandera, S. H. Henry Tye and J. Xu, JCAP {\bf 0705}, 004 (2007), hep-th/0702107;
X. Chen, Phys. Rev. {\bf D 72}, 123518 (2005), astro-ph/0507053; X. Chen, JHEP {\bf 0508}, 045 (2005), hep-th/0501184; X. Chen, Phys. Rev. {\bf D 71}, 063506 (2005), hep-th/0408084;
\bibitem{related}
G. Shiu and B. Underwood, Phys. Rev. Lett {\bf 98}, 051301 (2007), hep-th/0610151;
S. Kecskemeti, J. Maiden, G. Shiu and B. Underwood, JHEP {\bf 0609}, 076 (2006) hep-th/0605189;
H. V. Peiris, D. Baumann, B. Friedmann and A. Cooray, arXiv:0706.1240 [astro-ph];
X. Chen, S. Sarangi, S. H. Henry Tye and J. Xu, JCAP {\bf 0611}, 015 (2006), hep-th/0608082;
C. P. Burgess, J. M. Cline, K. Dasgupta and H. Firouzjahi, hep-th/0610320;
K. Dasgupta, H. Firouzjahi and R. Gwyn, hep-th/0702193;
H. Firouzjahi and S. H. Henry Tye, Phys. Lett. {\bf B 584}, 147-154 (2004);
F. Gmeiner and C. D. White, arXiv:0710.2009 [hep-th];
L. P. Chimento and R. Lazkoz, arXiv:0711.0712 [hep-th].
\bibitem{spinflation}
D. Easson, R. Gregory, G. Tasinato and I. Zavala, JHEP {\bf 0704}, 026 (2007), hep-th/0701252;
D. A. Easson, R. Gregory, D. F. Mota, G. Tasinato and I. Zavala, arXiv:0709.2666 [hep-th];
D. A. Easson, arXiv:0709.3757 [hep-th].
\bibitem{slowroll}
S. E. Shandera, JCAP {\bf 0504}, 011 (2005), hep-th/0412077;
M. Spalinski, JCAP {\bf 0704}, 018 (2007), hep-th/0702118.
\bibitem{nongaussianities}
M-x. Huang and G. Shiu, Phys. Rev. {\bf D 74}, 121301 (2006), hep-th/0610235;
X. Chen, M-x. Huang, S. Kachru and G. Shiu, JCAP {\bf 0701}, 002 (2007), hep-th/0605045;
D. Seery and J. E. Lidsey, JCAP {\bf 0701}, 008 (2007), astro-ph/0611034; J. E. Lidsey and D. Seery,
Phys. Rev. {\bf D 75}, 043505 (2007), astro-ph/0610398; D. Seery and J. E. Lidsey, JCAP {\bf 0509}, 011 (2005), astro-ph/0506056, D.Seery and J. E. Lidsey, JCAP {\bf 0506}, 003 (2005), astro-ph/0503692;
J. M. Maldacena, JHEP {\bf 0305}, 013 (2003), astro-ph/0210603..
\bibitem{kallosh}
R. Kallosh, N. Sivanandam and M. Soroush, arXiv:0710.3429 [hep-th];
T. W. Grimm, arXiv:0710.3833 [hep-th].
\bibitem{hawkingmoss}
S. W. Hawking and I. G. Moss, Phys. Lett. {\bf B 110}, 35 (1982).
\bibitem{kklt}
S. Kachru, R. Kallosh, A. Linde and S. P. Trivedi, Phys. Rev. {\bf D 68}, 046005 (2003), hep-th/0301240;
S. Kachru, R. Kallosh, A. Linde, J. Maldacena, L. McAllister and S. P. Trivedi, JCAP {\bf 0310}, 013 (2003), hep-th/0308055.
\bibitem{myers}
R. C. Myers, Class. Quant. Grav. {\bf 20}, S347-S372 (2003), hep-th/0303072;
R. C. Myers, JHEP {\bf 9912}, 022 (1999), hep-th/9910053.
\bibitem{tseytlin}
A. A. Tseytlin, hep-th/9908105;
A. A. Tseytlin, Nucl. Phys. {\bf B 501}, 41-52 (1997), hep-th/9701125.
\bibitem{klebanovstrassler}
I. R. Klebanov and M. J. Strassler, JHEP {\bf 0008}, 052 (2001), hep-th/0007191;
C. P. Herzog, I. R. Klebanov and P. Ouyang, hep-th/0108101.
\bibitem{finiten}
S. McNamara, C. Papageorgakis, S. Ramgoolam and B. Spence, JHEP {\bf 0605}, 060 (2006), hep-th/0512145;
C. Papageorgakis, S. Ramgoolam and N. Toumbas, JHEP {\bf 0601}, 030 (2006), hep-th/0510144.
S. Ramgoolam, B. J. Spence and S. Thomas, Nucl. Phys. {\bf B 603}, 237-276 (2004), hep-th/0405256.
\bibitem{giantinflaton}
O. DeWolfe, S. Kachru and H. Verlinde, JHEP {\bf 0405}, 017 (2004), hep-th/0403123;
S. Kachru, J. Pearson and H. Verlinde, JHEP {\bf 0206}, 021 (2002), hep-th/0112197.
\bibitem{jatkar}
D. P. Jatkar, G. Mandal, S. R. Wadia and K. P. Yogendran, JHEP {\bf 0201}, 039 (2002), hep-th/0110172.
\bibitem{n=1*}
C. Bachas, J. Hoppe and B. Pioline, JHEP {\bf 0107}, 041 (2001), hep-th/0007067.
\bibitem{reheating}
L. Kofman, A. Linde and A. Starobinsky, Phys. Rev. {\bf D 56}, 3258-3295 (1997), hep-ph/9704452;
L. Kofman and P. Yi, Phys. Rev. {\bf D 72}, 106001 (2005), hep-th/0507257:;
D. Chialva, G. Shiu and B. Underwood, JHEP {\bf 0601}, 014 (2006), hep-th/0508229;
J. H. Brodie and D. A. Easson, JCAP {\bf 0312}, 004 (2003), hep-th/0301138;
A. R. Frey, A. Mazumdar and R. Myers, Phys. Rev. {\bf D 73}, 026003 (2006);
C. Chen and S. H. Henry Tye, JCAP {\bf 0606}, 011 (2006), hep-th/0602136;
B. A. Bassett, S. Tsujikawa and D. Wands, Rev. Mod. Phys. {\bf 78}, 537-589 (2006), astro-ph/0507632.


\end{thebibliography}
\end{document}